\documentclass[journal]{IEEEtran}
\usepackage{graphicx}
\usepackage{subfigure}
\usepackage{multirow}
\usepackage[noend]{algpseudocode}
\usepackage{algorithmicx,algorithm}
\usepackage{dsfont}
\usepackage{amsfonts,amsmath}
\usepackage{url,cite}
\usepackage{color}
\def\cred{\textcolor{red}}  

\begin{document}

\title{Hyperspectral Unmixing via Deep Autoencoder Networks for a  Generalized Linear-Mixture/Nonlinear-Fluctuation Model}

\author{Min~Zhao,~\IEEEmembership{Student Member,~IEEE,}
        Mou~Wang,~\IEEEmembership{Student Member,~IEEE,}
        Jie~Chen,~\IEEEmembership{Senior Member,~IEEE,}
        and~Susanto~Rahardja,~\IEEEmembership{Fellow,~IEEE}
\thanks{The authors are with School of Marine Science and Technology, Northwestern Polytechnical University, China. (corresponding author: J. Chen, dr.jie.chen@ieee.org).}

\bigskip

\cred{\textbf{An improved version of this work, titled ``Hyperspectral Unmixing for Additive Nonlinear Models With a 3-D-CNN Autoencoder Network" has been published on \emph{IEEE Trans. Geosci. Remote Sens.} at \url{https://ieeexplore.ieee.org/abstract/document/9503107} (Open Access). }\\ Please refer to and cite the formal published version, which uses a 3DCNN to capture spatial correlation simultaneously. }
}

\markboth{IEEE xx,~Vol.~xx, No.~xx,~2019}%
{Shell \MakeLowercase{\textit{et al.}}: Hyperspectral Unmixing via Deep Autoencoder Networks for a  Generalized Linear-Mixture/Nonlinear-Fluctuation Model}

\maketitle

\begin{abstract}
Spectral unmixing is an important task in hyper- spectral image processing for separating the mixed spectral data pertaining to various materials observed aiming at analyzing the material components in observed pixels. Recently, nonlinear spectral unmixing has received particular attention as hyperspectral image processing, as there are many situations in which the linear mixture model may not be appropriate and could be advantageously replaced by a nonlinear one. Existing nonlinear unmixing approaches are often based on specific assumptions on the nonlinearity and can be less effective when used for scenes with unknown nonlinearity. This paper presents an unsupervised nonlinear spectral unmixing method that addresses a general model that consists of a linear mixture part and an additive nonlinear mixture part.  The structure of a deep autoencoder network, which has a clear physical interpretation, is specifically designed to achieve this purpose. The proposed scheme benefits from the universal modeling ability of deep neural networks and learning the nonlinear relation from the data. Extensive experiments with synthetic and real data, particularly with labeled laboratory-created data, illustrate the generality and effectiveness of this scheme compared with state-of-the-art methods.
\end{abstract}


\IEEEpeerreviewmaketitle

\section{Introduction}
\label{sec:intro}
\IEEEPARstart{H}{yperspectral} imaging is a continuously growing field of study that has received considerable attention over the past decade. Hyperspectral data provide high spectral resolution over a wide spectral range that typically extends from the infrared spectrum through the visible spectrum. This rich spectral information facilitates the discrimination of different materials in the observed scene. As a result, hyperspectral imaging has been widely adopted for a wide range of applications, such as land use analysis, pollution monitoring, wide-area reconnaissance and field surveillance~\cite{bioucas2012hyperspectral}.

However, the spectral content of individual pixels in hyperspectral images often represent a mixture of several materials from the imaged scene due to multiple factors, such as the low spatial resolution of hyperspectral imaging devices, the diversity of materials in the imaged scene and multiple reflections of photons from several objects. Therefore, separating spectra of individual pixels into a set of spectral signatures (endmembers), and determining the fraction abundances associated with each endmember is an essential task required for analyzing remotely sensed data. This process is denoted as spectral unmixing or mixed pixel decomposition~\cite{Keshava2002Spectral}. Spectral unmixing methods have been developed for this purpose based on both linear and nonlinear mixture models.

Among the presently available spectral mixture models, the linear mixture model (LMM) is the most widely used. In LMM, the incident light is assumed to be reflected by each component present in the scene only once prior to collection by the camera sensor, and the observed spectrum is thus a linear combination of the endmembers~\cite{Keshava2002Spectral}.
Many conventional unmixing methods based on LMM have been proposed. In \cite{yao2019nonconvex}, the total variation regularization is imposed to add spatial information of the hyperspectral image and \cite{hong2018augmented} considered the endmember variability problem.
While the LMM is simple and physically interpretable, numerous complex conditions arise where the incoming light may undergo complex interactions among the individual materials in the scene, resulting in higher-order photon interactions that introduce nonlinear effects in the mixed spectra. Consequently, the analysis of data collected under these conditions requires nonlinear unmixing methods~\cite{Heylen2014A}. A considerable number of studies have recently focused on addressing  nonlinear unmixing problems. For example, bilinear models~\cite{Borel1994Nonlinear}  have been developed to address conditions of  second-order scattering interactions that may occur on complex vegetated surfaces, by adding extra bilinear interaction terms to the linearly composited spectrum. Such models include the Fan model~\cite{Fan2009Comparative} and the generalized bilinear model (GBM)~\cite{Halimi2011Unmixing}. The polynomial post-nonlinear mixture model (PNMM) applies a polynomial function to the linearly mixed data to approximate the nonlinearity of photon interactions occurring in an imaged scene~\cite{Altmann2012Supervised}. A bidirectional reflectance model has been developed to describe the photon interactions of intimately mixed particles based on the fundamental principles of radiative transfer theory. This model is generally referred to as the intimate mixture model or Hapke model~\cite{Hapke1981Bidirectional}. The multimixture pixel (MMP) model further extended the intimate mixture model by integrating it with the LMM model~\cite{Close2012Using}. The above cases have been generalized by considering a linear-mixture/nonlinear-fluctuation (K-Hype) model, where the nonlinear fluctuation was described by a function defined in a reproducing kernel Hilbert space (RKHS)~\cite{chen2013nonlinear}.  Further extensions of this model have also been proposed with spatial regularization~\cite{Chen2014TGRS} and neighborhood-dependent contributions~\cite{ammanouil2017nonlinear}. The multilinear mixing model (MLM) considers an infinite number of photon interactions by introducing a probability of photon undergoing further interactions~\cite{heylen2016multilinear}.
The work \cite{heylen2019nonlinear} uses a graph-based model to describe the multiple photon interactions.
However, most of the above models rely heavily on specific assumptions regarding the inherent nonlinearity of the spectral unmixing, and they are therefore not well suited to scenes with unknown nonlinearity characteristics. In addition, while the K-Hype model based on the RKHS presented above and other kernel-based algorithms provide flexible nonlinear modeling, the selection of appropriate kernels and kernel parameters has been demonstrated to be a non-trivial issue that restricts the application of these approaches. Finally, all of these algorithms assume that the endmembers are known prior, and therefore focus strictly on evaluating the abundance fractions.

In recent years, deep learning has demonstrated its superior performance in addressing various nonlinear problems compared to classical methods. Researchers have also investigated the use of deep neural networks in hyperspectral image analysis. Particular attention has been focused on the hyperspectral image classification problem~\cite{chen2014deep,mou2017deep,chen2015spectral}. However, despite the recognized potential of neural networks for solving inverse problems, only a handful of studies have applied neural networks for addressing the spectral unmixing problem. Among these, classifier models have been applied to spectral unmixing~\cite{licciardi2011pixel,zhang2018hyperspectral}, but this approach requires a training set with known ground-truths, which must often be generated by theoretical models. In addition, autoencoder networks have also been applied to the blind spectral unmixing problem. An autoencoder is a network that learns to compress an input into a short code which can be uncompressed into something that is close to the original input. Internally, it has a hidden layer that describes that short dimensional code used to represent the input data for reconstructing the output data. This is ideally suited for conducting spectral unmixing because this process can also be considered as finding a low dimensional representation (abundance fractions) of hyperspectral data. For example, approaches employing autoencoder networks have exhibited good performance in determining both endmembers and abundance fractions\cite{1guo2015hyperspectral,1palsson2018hyperspectral,1qu2017spectral,1qu2018udas,1su2017nonnegative,1su2018stacked,ozkan2018endnet,borsoi2019deep}.

However, these approaches are specifically designed to preprocess the input data or address the linear unmixing problem, and therefore fail to make use of the superior potential of neural networks for addressing nonlinear problems, while linear unmixing is readily addressed using classical methods. These issues were addressed in our previous work~\cite{NAE2019}, where we designed an autoencoder network for conducting blind nonlinear unmixing. This work considered a post-nonlinear spectral mixture, where the post-nonlinearity was modeled by the decoder part of the autoencoder.  In there,  pretraining and learning rate adjustment techniques were required to ensure the effectiveness of the decoder, and the nonlinear model represented by the decoder was not sufficiently general to cover multiple nonlinear cases.

The present work addresses the deficiencies in previous works by re-examining the nonlinear mixture models and restrictions of existing unmixing schemes based on deep neural networks. Accordingly, this paper presents a new autoencoder network structure for blind nonlinear spectral unmixing.  
The highlights of this work are summarized as follows.
\begin{enumerate}
  \item A general  spectral mixture model that consists of a linear mixture component and an additive nonlinear mixture component is proposed.   The significance of an endmember in the nonlinear mixture component is weighted according to its associated abundance fraction. A deep neural network is proposed to represent this nonlinear part and generalizes the existing related models.
  \item A deep autoencoder network is designed to conduct the nonlinear unmixing based on the proposed model. The form of the inherent nonlinearity of the nonlinear mixture component is learned from the data itself, rather than relying on an assumed form. The structure of the decoder is designed with particular care so that the nonlinear interactions are imposed on endmembers weighted by abundances, which has a clear physical interpretation and covers several existing artificial models. Endmembers and abundance fractions are extracted from the outputs and weights of the particular layers of the network. Extra regularizations are also imposed to enhance the unmixing performance.
  \item Most existing studies have relied on numerically produced synthetic data and an intuitive inspection of the results of real data. Lack of publicly available datasets with ground-truths hampers capability to evaluate and compare the performance of unmixing algorithms in a quantitative and objective manner. The proposed algorithm is tested using real data with ground-truths created in our laboratory. Using labeled real data provides for more convincing comparison results.
\end{enumerate}

The remainder of this paper is organized as follows: Section~II presents the formulation of the nonlinear mixture model. Section~III presents the design of the proposed autoencoder scheme for unmixing.  Section~IV validates the proposed method with experiments using synthetic and real data. Section~V concludes the work and provides the perspective of the future work.

\begin{table*}[!t]
\centering
\caption{\small Relating typical nonlinear models with the generic form Eq.~\eqref{eq.NMs} (noise vector $\mathbf{n}$ is omitted for saving space).}
\label{tab:relation}
\begin{tabular}{c|l|l|l}
\hline\hline
                         &  \multicolumn{1}{c|}{Model expression}                  &  \multicolumn{1}{c|}{Form of $\Psi$}  &  \multicolumn{1}{c}{Note}\\
\hline
Bilinear model           & $\mathbf{x}=\mathbf{M}\mathbf{a}+ \sum_{i=1}^R\sum_{j=i+1}^R a_i\mathbf{m}_i \odot a_j\mathbf{m}_j $ &     $\Psi = \sum_{i=1}^R\sum_{j=i+1}^R a_i\mathbf{m}_i \odot a_j\mathbf{m}_j$ & $\odot$ denotes the element-wise product  \\
\hline
Post-nonlinear model     & $\mathbf{x}=\mathbf{M}\mathbf{a}+ \mathbf{M}\mathbf{a}\odot \mathbf{M}\mathbf{a} $       & $\Psi = \mathbf{M}\mathbf{a}\odot \mathbf{M}\mathbf{a}$ &   Note $\mathbf{M}\mathbf{a} = \mathbf{m}_1 a_1+\cdots + \mathbf{m}_R a_R$    \\
\hline
K-Hype model             &    $x_i=\mathbf{m}_i\mathbf{a}+ \psi(\mathbf{m}_{\lambda_i})$ & $[\Psi]_i = \sum_{i=1}^B \beta_i \,\kappa(\mathbf{m}_{\lambda_i}, \mathbf{m}_{\lambda_j})$  & \hspace{-3mm} $\begin{array}{l} \psi \text{ is in a RKHS with kernel } \kappa \\ \beta_i \text{ are coefficients to be determined} \\ \mathbf{a} \text{ is ignored in } \Psi\end{array}$\\
\hline
Multilinear mixing model & $\mathbf{x}=\mathbf{Ma}+p\mathbf{Ma}(\mathbf{Ma}-1)/(1-p\mathbf{Ma})$                                                                                                                 &  $\Psi =  p\mathbf{Ma}(\mathbf{Ma}-1)/(1-p\mathbf{Ma})$  & $p$ is the probability of interactions\\
\hline\hline
\end{tabular}
\end{table*}

\section{Problem formulation}

\noindent\textbf{Notation.} Normal font $x$ and $X$ denote scalars. Boldface small letters $\mathbf{x}$ denote vectors. All vectors are column vectors. Boldface capital letters $\mathbf{X}$ denote matrices. Considering an observed pixel data $\mathbf{x}\in \mathbb{R}^{B}$ with $B$ denoting the number of spectral bands, and $\mathbf{M} = [\mathbf{m}_1, \cdots, \mathbf{m}_R]$ denotes the $(B\times R)$ endmember matrix with endmembers $\mathbf{m}_i$, $R$ represents the number of endmembers. $\mathbf{a}=[a_1, a_2, \cdots, a_R]^\top$ is the abundance vector associated with a pixel. The operator $\text{blkdiag}\{\cdots\}$ forms a matrix of size $BR\times R$ using vectors $\{\mathbf{y}_i\}_{i=1}^R \in\mathbb{R}^B$ such that
\begin{equation}
\text{blkdiag}\{\mathbf{y}_1, \mathbf{y}_2, \cdots, \mathbf{y}_R\} =
\begin{bmatrix} \mathbf{y}_1 & \mathbf{0}_B  &  \cdots &  \mathbf{0}_B \\
                         \mathbf{0}_B   & \mathbf{y}_2 & \cdots &  \mathbf{0}_B  \\
                          \vdots        &    \vdots         &   \vdots &  \vdots       \\
                          \mathbf{0}_B  & \mathbf{0}_B & \cdots& \mathbf{y}_R \end{bmatrix} \in \mathbb{R}^{BR\times R}
\end{equation}
with $\mathbf{0}_B$ denoting all zero vectors of length $B$. Considering using such a matrix $\mathbf{Y} =\text{blkdiag}\{\mathbf{y}_1, $ $\mathbf{y}_2, \cdots, \mathbf{y}_R\}$  as the weight matrix of a layer of a deep neural network, regular matrix product maps the input $\mathbf{h}=[h_1, \cdots, h_R]^\top$ to the output of the form
\begin{equation}
\label{eq:blkdiagprod}
         \mathbf{Y}\mathbf{h} = \big[h_1 \mathbf{y}_1^\top , h_2\mathbf{y}_2^\top, \cdots,h_R\mathbf{y}_R^\top\big]^\top.
\end{equation}
The usefulness of such an operation will be clear when we relate Eq.~\eqref{eq:Vh} and Eq.~\eqref{eq:recx} to Eq.~\eqref{eq.our_model}.

The operator $\text{col}\{\mathbf{y}_1, \cdots,\mathbf{y}_N\}$ stacks its vector arguments $\{\mathbf{y}_i\}_{i=1}^N$ on the top of each other to generate a connected vector given by
\begin{equation}
        \text{col}\{\mathbf{y}_1, \cdots, \mathbf{y}_N\} =[\mathbf{y}_1^\top, \cdots, \mathbf{y}_N^\top]^\top=\left(\begin{array} {c} \mathbf{y}_1 \\ \vdots \\ \mathbf{y}_N \end{array}\right).
\end{equation}
\medskip

We firstly consider the linear mixing model where each observed pixel is assumed to be a linear combination of the endmembers weighted by their associated abundances:
\begin{equation}\label{eq.LMM}
  \mathbf{x}=\mathbf{M}\mathbf{a}+\mathbf{n},
\end{equation}
where $\mathbf{n}\in \mathbb{R}^B$ is an additive noise vector. As the abundances represent relative fractions of each material, they are required to satisfy the abundance non-negative constraint (ANC), Eq.~\eqref{eq.ANC}, and abundance sum-to-one constraint (ASC), Eq.~\eqref{eq.ASC}, that are
\begin{equation}\label{eq.ANC}
\forall i: a_i\geq 0
\end{equation}
\begin{equation}\label{eq.ASC}
  \sum_{i=1}^{R}a_i=1\cred{.}
\end{equation}
In this work, we consider the following general mixing mechanism:
\begin{equation}\label{eq.NM}
  \mathbf{x}=\mathbf{M}\mathbf{a}+\Psi(\mathbf{M},\mathbf{a})+\mathbf{n},
\end{equation}
which consists of a linear mixture of endmembers $\mathbf{M}$ with abundance fractions $\mathbf{a}$, and a nonlinear fluctuation $\Psi$ that defines the interactions of $\mathbf{M}$ parameterized by $\mathbf{a}$.
Several existing typical nonlinear models are summarized in Table~I.  We revised the mixture model defined in Eq.~\eqref{eq.NM} to provide a more tractable form as follows:
\begin{equation}\label{eq.NMs}
    \begin{split}
         \mathbf{x}&=\mathbf{M}\mathbf{a}+\Psi(a_1\mathbf{m}_1, a_2\mathbf{m}_2, \cdots, a_R\mathbf{m}_R)+\mathbf{n} \\
                         &=\mathbf{M}\mathbf{a}+\Psi(\mathbf{M}\,\text{diag}(\mathbf{a}))+\mathbf{n}.
  \end{split}
\end{equation}
where $\Psi$ represents the nonlinear interaction between  the endmembers $\mathbf{M}$ weighted by associated abundance $\mathbf{a}$. It is clear that
Eq.~\eqref{eq.NMs} is a general model that yields the models defined in Table I under different choices of $\Psi$.
We refer to this model as a \emph{gerenalized linear-mixture/nonlinear-fluctuation model}. This form suggests that the nonlinear interactions of material signatures are in proportion to the abundance fractions of each material. This is reasonable, because, for instance,  a material with a negligible abundance will have limited contribution to either the linear component, or the nonlinear component of $\mathbf{x}$. Several existing nonlinear models can be considered as  specific cases of Eq.~\eqref{eq.NMs} under different definitions of $\Psi$. Typical nonlinear mixing models and the relations between these algorithms and Eq.~\eqref{eq.NMs} are summarized in Table~\ref{tab:relation}. With the exception of K-Hype, these algorithms are designed manually to capture the assumed nonlinearities.  The linear-mixture/nonlinear-fluctuation model used by the K-Hype algorithm is relatively more general, and has some similarities with Eq.~\eqref{eq.NMs}. However, in addition to the non-trivial issue associated with the selections of kernel and kernel parameters discussed above, this model suffers from the use of a nonlinear fluctuation function that is independent of the abundance fractions. Hence, the endmembers contribute equivalently to the nonlinear component of the observed spectrum.

The present model clearly addresses this restriction by explicitly including the abundance fractions of the endmembers. In addition, the restriction associated with the selections of kernels and kernel parameters is addressed in the proposed approach by not assigning $\Psi$ in Eq.~\eqref{eq.NMs} with any specific form. Instead, we devise a method to learn it from the data itself via an autoencoder network.


In order to facilitate to present the structure of the autoencoder network, we write Eq.~\eqref{eq.NM} in the following equivalent form
\begin{equation}\label{eq.our_model}
  \mathbf{x} = \mathcal{T}(\mathbf{M}_D\,\mathbf{a})+ \Psi(\mathbf{M}_D\,\mathbf{a})+\mathbf{n}
\end{equation}
where
\begin{equation}
      \mathbf{M}_D=\text{blkdiag}\{\mathbf{m}_1, \mathbf{m}_2, \cdots, \mathbf{m}_R \}\in \mathbb{R}^{BR\times R}
\end{equation}
and $\mathcal{T}: \mathbb{R}^{BR} \mapsto \mathbb{R}^B$ is a step-wise summation operator, i.e., for a given vector $\mathbf{y} \in\mathbb{R}^{BR}$
\begin{equation}
            \label{eq.our_model2}
            \mathcal{T}(\mathbf{y}) = \left( \sum_{i=1}^R y_{B \times (i-1)+1}, \cdots, \sum_{i=1}^R y_{B \times (i-1)+B} \right)^\top.
\end{equation}
With the above notation, we have
\begin{equation}
      \mathbf{M}_{D}\mathbf{a}=\text{col}\{a_{1}\mathbf{m}_1, a_{2}\mathbf{m}_2, \cdots, a_{R}\mathbf{m}_R\},
\end{equation}
and
\begin{equation}
      \label{eq.equMa}
      \mathcal{T}(\mathbf{M}_D\mathbf{a})=\mathbf{M}\mathbf{a}.
\end{equation}
Then the linear component and the nonlinear component shares the same input $\mathbf{M}_D\mathbf{a}$.

\section{Proposed Approach}
In this section, we present a thorough presentation of the proposed  method that solves the nonlinear unmixing problem using deep autoencoder networks.

\subsection{General structure}
\begin{figure*}[!t]
  \centering
  \centerline{\includegraphics[width=19cm]{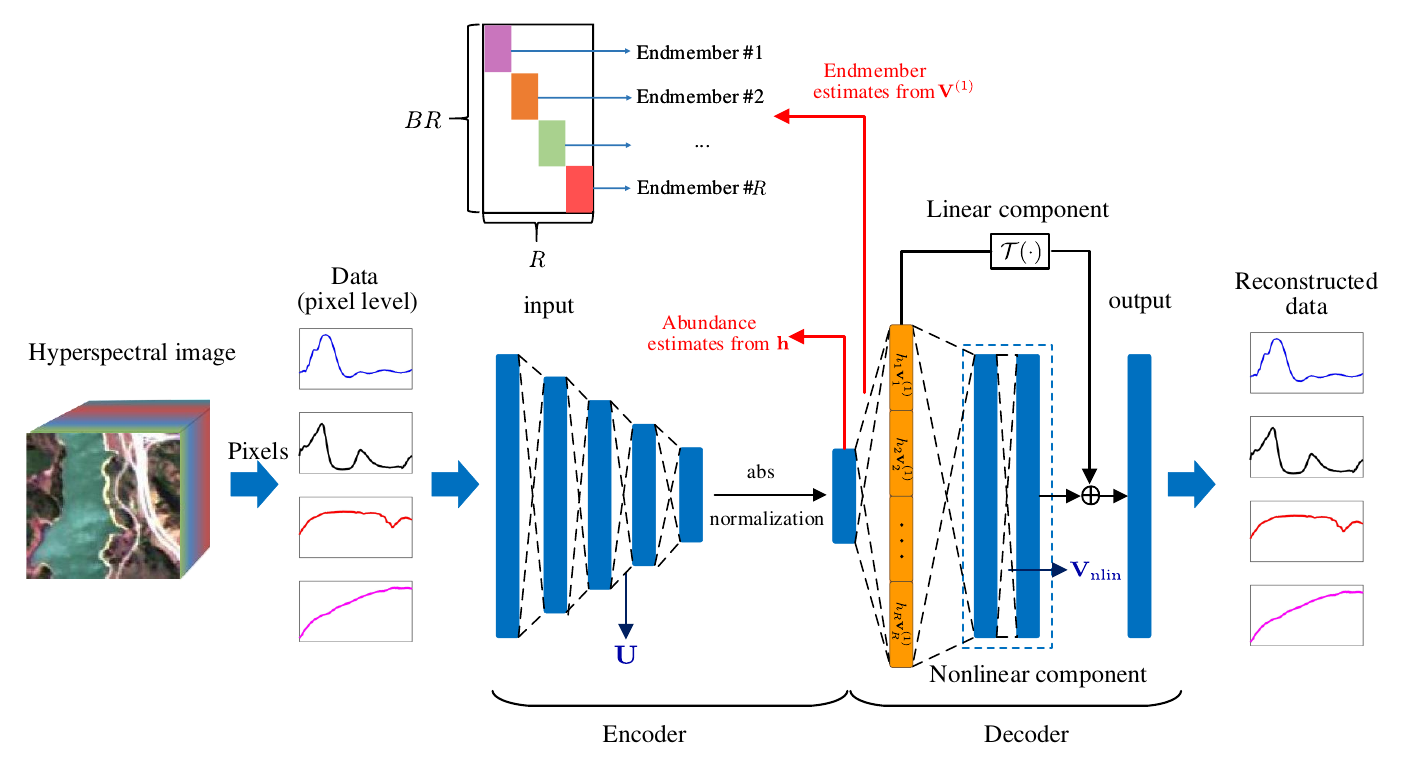}}
  \vspace{-3mm}
  \caption{Diagram of the proposed system. As presented in Sec. III, $\mathbf{U}$ are the weights of the encoder, $\mathbf{h}$ is the output of the utility layer (Eq.~\eqref{eq.sum_one}), $\mathbf{V}^{(1)}$ are the weights of the first-layer of the decoder, and $\mathbf{V}_{\rm nlin}$ denotes the weights of the nonlinear part of the decoder.}
\label{fig.framework}
\vspace{-5mm}
\end{figure*}

The structure of an autoencoder network consists of two parts, namely an encoder and a decoder. Encoder $f_{\rm E}$ compresses the input $\mathbf{x}$ into a low dimensional representation $\mathbf{h}\in\mathbb{R}^R$, i.e.
\begin{equation}
           \mathbf{h}=f_{\rm E}(\mathbf{x}),
\end{equation}
with $f_{E}: \mathbb{R}^{B\times 1}\rightarrow \mathbb{R}^{R\times 1}$.
Recall that $B$ represents the number of bands and $R$ denotes the number of endmembers.
Decoder $f_{\rm D}$ uncompresses the hidden representation vector $\mathbf{h}$ to reconstruct the original input data, i.e.,
\begin{equation}
        \hat{\mathbf{x}}=f_{\rm D}(\mathbf{h}),
\end{equation}
with $f_{\rm D}: \mathbb{R}^{R\times 1}\rightarrow \mathbb{R}^{B\times 1}$.
The network trains the parameters and representations by minimizing the average reconstruction error between the input $\mathbf{x}$ and its reconstructed counterpart $\mathbf{\hat{x}}_i=f_{\rm D}(f_{\rm E}(\mathbf{x}_i))$ given by
\begin{equation}\label{eq.loss_error}
 \mathcal{L}(\mathbf{x}, \hat{\mathbf{x}})=\frac{1}{N}\sum_{i=1}^{N}\big\|\hat{\mathbf{x}}_{i}-\mathbf{x}_{i}\big\|^{2}.
\end{equation}
With the output of the encoder $\mathbf{h}\in\mathbb{R}^R$, in this work the decoder is designed to reconstruct the input $\mathbf{x}$ with the following specific structure:
\begin{equation}
         \label{eq.decoder}
         \hat{\mathbf{x}} = \mathcal{T}(\mathbf{V}^{(1)} \,\mathbf{h}) + \Phi( \mathbf{V}^{(1)} \,\mathbf{h})
\end{equation}
where $\mathbf{V}^{(1)}$ are weights of the first layer of the decoder, as to be defined in Eq.~\eqref{eq.defV1}, $\Phi$ is the nonlinear function constructed by the nonlinear part of our decoder, and It is expected that after the learning process, $\mathbf{\Phi}$ mimics the generative model~$\mathbf{\Psi}$. Comparing this structure to Eq.~\eqref{eq.our_model}, the decoder mimics the output in accordance to this model. Therefore, after the network parameters are learnt with data, blind unmixing of the same input data can be conducted by:
\begin{eqnarray}
  \mbox{Abundance estimation} &:& \mathbf{h}\Rightarrow \hat{\mathbf{a}} \\
  \mbox{Endmember extraction} &:& \mathbf{V}^{(1)}\Rightarrow \widehat{\mathbf{M}}_D.
\end{eqnarray}
Both encoder and decoder can either be shallow or deep, but generally, it is believed that deep networks possess a superior modeling capability. The schema of the proposed autoencoder network is illustrated in Figure~\ref{fig.framework} in order that readers can better understand the proposed structure.
We elaborate the design of encoder and decoder in the following subsections.

\subsection{Encoder}
In this work, a regular deep network is designed as the encoder with a useful specific structure reported in the upper-part of Table~\ref{Tab.network_structure}.
No specific constraints are imposed on the encoder in order to fully use the capacity of the network and reduce the information loss. The number of units of the input layer is the same as the number of spectral bands $B$,  and the number of units of the last layer is the number of endmembers $R$.
Note that the number of endmembers $R$ is assumed priorly known in our work.
The four fully connected layers gradually narrow down from the input layer of dimension $B$ to deep layers until reaching the size $R$. Except for the last hidden layer, the first three layers adopt the same activation functions $\phi$, such as  Sigmoid, ReLU, and Leaky ReLU (LReLU).
ReLU is one of the most notable activation function in modern deep learning systems, and the LReLU is considered an improved version as it has nonzero gradient for all inputs. We have conducted extensive experiments to validate the enhanced performance of LReLU. Hence, LReLU is preferred in this work.
The relation between the input and the output of layer $i$ of the encoder is given by
\begin{equation}\label{eq.ac_function}
  \mathbf{h}^{(i)}=\phi\big(\mathbf{U}^{(i)}\mathbf{h}^{(i-1)}+\mathbf{b}^{(i)}\big),
\end{equation}
where  $\mathbf{h}^{(i-1)}$ and $\mathbf{h}^{(i)}$ represent the outputs of the previous layer and the current layer respectively, $\mathbf{U}^{(i)}$ and $\mathbf{b}^{(i)}$ are the weight matrix and the bias vector of the current layer.

The non-negativity and sum-to-one constraints imposed on abundance vector $\mathbf{a}$ should be carefully addressed. In order to meet the ANC, the work~\cite{1qu2018udas} uses a threshold to enforce the vector to be non-negative, and the work~\cite{1su2018stacked} uses a non-negative autoencoder to guarantee the ANC over the whole network. The former strategy deactivates a large number of nodes in the network, and the capability of network is thus not fully utilized. The latter strategy imposes strong constraints on the network and makes it difficult to design the network.  For the ASC, the works~\cite{1qu2018udas} and \cite{1su2018stacked} add a regularization to encourage the ASC, and \cite{1palsson2018hyperspectral} uses a normalization operator on $\mathbf{a}$.  In this work, we address the ANC and ASC using the strategy proposed in our previous work~\cite{NAE2019}. Absolute value rectification is used to enforce $\mathbf{h}$, the output of the encoder network (abundance estimation), to be non-negative. Then this non-negative vector is normalized by sum of its entries to satisfy sum-to-one, namely,
\begin{equation}\label{eq.sum_one}
  h_{i}=\frac{|h_i|}{\sum_{i=1}^{R}|h_i|},
\end{equation}
where $h_{i}$ is the $i$th element of the abundance vector $\mathbf{h}$.
{\begin{table}[!t]
\footnotesize
\centering
\renewcommand{\arraystretch}{1.4}
\caption{The structure of network. (``Utility'' denotes a layer that performs some transforms other than the regular activations on the outputs of the previous layer.)}
\label{Tab.network_structure}
\begin{tabular} {|c|c|c|c|c|}
\hline
\multirow{6}{*}{Encoder}   & \multicolumn{2}{c|}{Layers}                & Activation function  & unit   \\ \cline{2-5}
                           & \multicolumn{2}{c|}{Input layer}                   & -                    & $B$   \\
                           & \multicolumn{2}{c|}{Hidden layer}                  & LReLU             & $32R$    \\
                           & \multicolumn{2}{c|}{Hidden layer}                  & LReLU            & $16R$     \\
                           & \multicolumn{2}{c|}{Hidden layer}                  & LReLU               & $4R$     \\
                           & \multicolumn{2}{c|}{Hidden layer}                  & -                    & $R$      \\ \hline
\multirow{1}{*}{Utility}   & \multicolumn{2}{c|}{abs + normalization}           & -                    & $R$    \\ \hline
\multirow{3}{*}{Decoder}   & Linear part                        & Hidden layer  & ReLU               & $BR$    \\ \cline{2-5}
                           & \multirow{2}{*}{Nonlinear part}    & Hidden layer  & LReLU              & $B$    \\
                           &                                    & Hidden layer  & LReLU              & $B$    \\
                           &                                    & Output layer  & ReLU                    & $B$    \\
\hline
\end{tabular}
\vspace{-3mm}
\end{table}}

\subsection{Decoder}

The decoder is designed to reconstruct the input with a linear structure and a parallel nonlinear structure. The specific setting of this structure is reported in the lower-part of Table~\ref{Tab.network_structure}. Recalling the operators and symbols defined in Eq.~\eqref{eq.our_model} to Eq.~\eqref{eq.equMa} and the decoder structure given by Eq.~\eqref{eq.decoder}, the first layer of the decoder is then designed by
\begin{equation}\label{eq.decoder_linear}
  \mathbf{o}^{(1)}= \mathbf{V}^{(1)}\mathbf{h},
\end{equation}
where $\mathbf{V}^{(1)}$ is defined as weights of the first layer of the decoder. 
Endmember extracted by linear algorithms, like the VCA algorithm can be used to initialize the learning process of $\mathbf{V}^{(1)}$.
$\mathbf{V}^{(1)}$ is constrained with the following form
\begin{equation}
       \label{eq.defV1}
       \mathbf{V}^{(1)} = \text{blkdiag}\{\mathbf{v}^{(1)}_1, \cdots, \mathbf{v}^{(1)}_R\}.
\end{equation}
Consequently, the product $\mathbf{V}^{(1)} \mathbf{h}$  equals to
\begin{equation}
         \label{eq:Vh}
         \mathbf{V}^{(1)} \mathbf{h} = \text{col}\{h_1\mathbf{v}^{(1)}_1, \cdots, h_R\mathbf{v}^{(1)}_R \}.
\end{equation}
Vectors $\{h_i\mathbf{v}_i^{(1)}\}_{i=1}^R$ are generated as estimates of the endmembers weighted by the associated abundances.
 The output $\mathbf{o}^{(1)}$  of this layer is used as the input of $\mathcal{T}(\cdot)$ so that $\mathcal{T}(\mathbf{o}^{(1)})$ generates the linear component of the spectrum, and $\mathbf{o}^{(1)}$  is also used as the input of the nonlinear component defined by a fully connected network without bias weights. This nonlinear component of decoder is designed to represent the nonlinear interactions among the endmembers weighted by the associated abundances.
Studies show that a neural network with two hidden layers can represent arbitrary nonlinear relation among the input~\cite{goodfellow2016deep}.
In our scheme, we use two hidden layers to learn the nonlinear relation among the endmembers, since very high-order photon interactions, though may exist, are usually weak in practice.
To avoid over-fitting, a parameter norm penalty is added to the weights of the nonlinear component. We shall elaborate this parameter penalty in the next subsection.   This network thus learns the nonlinearity from the data and models all nonlinear interactions among $\{h_i\mathbf{v}^{(i)}_i\}_{i=1}^R$. Finally, the outputs of these two parallel structures are added to reconstruct the estimate $\hat{\mathbf{x}}$ by:
\begin{eqnarray}\label{eq.out_decoeder}
  \hat{\mathbf{x}} &=& f_{\rm D}(\mathbf{h}) \\
                   &=& \hat{\mathbf{x}}_{\text{lin}}+\hat{\mathbf{x}}_{\text{nlin}} \\
                   &=& \mathcal{T}(\mathbf{o}^{(1)})+\Phi(\mathbf{o}^{(1)}).  \label{eq:recx}
\end{eqnarray}
The energy of component $\hat{\mathbf{x}}_{\text{nlin}}$ allows to indicate where the nonlinear effects spatially appear, which can be useful in many applications.

\subsection{Objective function}
Several components are considered to formulate the objective function of the proposed autoencoder. The mean-square error between the input and reconstructed data is employed for the data fitting:
\begin{equation}\label{eq.mse}
 J_{\rm data}(\mathbf{W})=\mathcal{L}(\mathbf{x}, \mathbf{\hat{x}})=\frac{1}{N}\sum_{i=1}^{N}\|f_{\rm D}(f_{\rm E}(\mathbf{x}_i))-\mathbf{x}_{i}\|^{2}.
\end{equation}
Blind unmixing problem with both endmember and abundance unknown can be a difficult inverse problem. Regularization is often imposed to condition the problem with reasonable prior information. In this work, we first consider the regularity of the nonlinear function $\Psi$, as proposed in~\cite{chen2013nonlinear}.  Thus the $\ell_2$-norm of the weights  of the nonlinear part of the decoder (denoted by $\mathbf{V_{\rm nlin}}$) given by
\begin{equation}
      J_{\rm reg}(\mathbf{V}_{\rm nlin}) = \|\mathbf{V_{\rm nlin}}\|^2
\end{equation}
is used as the regularization to drive the weights to decay and avoid over-fitting.  Further, a first-order TV-norm regularization given by
\begin{equation}
     J_{\rm smth} (\mathbf{V}^{(1)}) = \sum_{i=1}^R\sum_{j=1}^{B-1}\left| [\mathbf{v}^{(1)}_i]_{j+1} - [\mathbf{v}^{(1)}_i]_{j} \right|
\end{equation}
is imposed on $\{\mathbf{v}^{(1)}_i\}_{i=1}^R$.  Because $\{\mathbf{v}^{(1)}_i\}_{i=1}^R$ are the estimates of the endmembers, such a regularization encourages the smoothness of the endmembers and reduces the estimation noise. Finally, the objective function is formulated by
\begin{equation}
      \label{eq.obj}
       J(\mathbf{W}) =  J_{\rm data}(\mathbf{W}) + \lambda\,  J_{\rm reg}(\mathbf{V}_{\rm nlin}) + \gamma\, J_{\rm smth} (\mathbf{V}^{(1)})
\end{equation}
where positive parameters $\lambda$ and $\gamma$ control the strengths of the two regularization terms.

\section{Experiments}
In this section, the proposed unmixing scheme was implemented and its performance was compared with several typical state-of-the-art unmixing methods, using  synthetic data, labeled laboratory-created data, and real airborne image data. Note that the general network structure and number of layers are the same for all experiments and all data are normalized between 0 and 1.

The performance of abundance estimation was measured by the root mean square error (RMSE) defined by
\begin{equation}\label{eq.rmse}
  {\rm RMSE} = \sqrt{\frac{1}{NR}\sum_{i=1}^{N}\big\|\mathbf{a}_{i}-\hat{\mathbf{a}}_{i}\big\|^{2}}
\end{equation}
where $N$ represents the number of pixels, $\mathbf{a}_{i}$ and $\hat{\mathbf{a}}_{i}$ denote the true and estimated abundance vectors of the $i$th pixel.

The accuracy of the endmember estimation was evaluated using the spectral angle distance (SAD) and the spectral information divergence (SID) given by
\begin{equation}
\left\{
             \begin{array}{lr}
             {\rm SAD} = \cos^{-1}\left(\frac{\mathbf{m}^{T}\widehat{\mathbf{m}}}{\|\mathbf{m}\|\|\widehat{\mathbf{m}}\|}\right) &  \\
             {\rm SID}(\mathbf{m}|\widehat{\mathbf{m}})=\sum_{j}\mathbf{p}_{j}\log\left(\frac{\mathbf{p}_{j}}{\hat{\mathbf{p}}_{j}}\right), &
             \end{array}
\right.
\end{equation}
where $\mathbf{m}$ represents an endmember and $\widehat{\mathbf{m}}$ represents its estimate,  $\mathbf{p}=\frac{\mathbf{m}}{\mathbf{1}^{\top}\mathbf{m}}$ is the probability distribution vector of each endmember, and $\hat{\mathbf{p}}=\frac{\hat{\mathbf{m}}}{\mathbf{1}^{\top}\hat{\mathbf{m}}}$.

The following typical algorithms were compared:
\begin{itemize}
  \item \textbf{The endmember extraction with VCA and abundance estimation with K-Hype~\cite{chen2013nonlinear}:} VCA is a classic geometric method used for endmember extraction. The K-Hype algorithm considers the linear-mixture/nonlinear-fluctuation model and approximates the nonlinearity by the kernel trick.
  \item \textbf{The endmember extraction with VCA and abundance estimation with multilinear model (MLM)~\cite{heylen2016multilinear}:} MLM is based on a Markov chain interpretation of the reflection process of a single light ray. A probability parameter is used to describe the possibility of interacting with the next material.
  \item \textbf{The endmember extraction with N-FINDR and abundance estimation with NDU~\cite{ammanouil2017nonlinear}} N-FINDR \cite{winter1999n} is a classic method that used to extract endmember. NDU is a nonlinear abundance estimation method that is band-dependent and uses neighborhood information.
  \item \textbf{The robust non-negative matrix factorization (rNMF)~\cite{F2015Nonlinear}:} rNMF is an NMF-based nonlinear method that determines the endmembers and abundances simultaneously via a block-coordinate descent algorithm that involves majorization-minimization updates.
  \item \textbf{A deep autoencoder network for nonlinear unmixing (NAE)~\cite{NAE2019}:} NAE is a novel scheme for blind nonlinear unmixing based on a deep autoencoder network that addresses the post-nonlinear mixture problem.
  \item \textbf{NUSAL~\cite{halimi2016fast}:} it is a kernel based method for nonlinear unmixing by variable splitting and augmented Lagrangian. The method also assumes a linear mixing model corrupted by an additive term whose expression can be adapted to account for multiple scattering nonlinearities.
  \item \textbf{SAE~\cite{xu2018supervised}:} it is a supervised abundance estimation method for nonlinear unmixing based on a classifier model. This method uses stacked autoencoder scheme to learn the mapping between pixels spectra and the fractional abundances.
   \end{itemize}
Note that the linear endmember extraction algorithms were used for the first three methods that are focused on the abundance estimation. These geometrical algorithms still provide sufficiently good results when the nonlinearity degree in data is moderate, as they are able to extract vertices from distorted data clouds~\cite{dobigeon2014nonlinear}. Our experiments will also confirm their performance. All unsupervised nonlinear unmixing methods, namely rNMF, NAE and our proposed method, are initialized by the same VCA result.

\begin{figure*}[!t]
  \centering
  \centerline{\includegraphics[width=18cm]{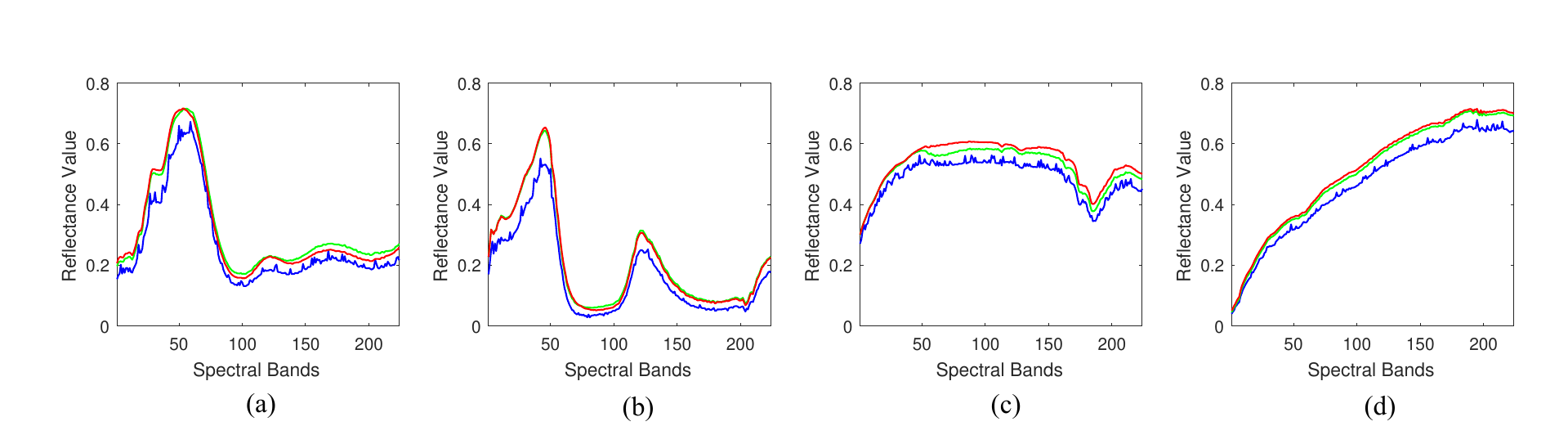}}
  \caption{Illustration of extracted four endmembers from the data with the linear model mixture under SNR=20 dB. Red curves represent the ground-truth. The blue and green curves represent the extracted endmembers with $\gamma=0$ and $\gamma=10^{-3}$ respectively. Proper regularization increases the smoothness of the estimated endmembers.}
\label{fig.curves}
\end{figure*}

\begin{table*} [!t]
\small \centering
\caption{\small Abundance RMSE Comparison of the synthetic data.}
\vspace{-1mm}
  \begin{tabular}{c|c|c|c|c|c|c|c|c|c}
     \hline
     \hline
                &\multicolumn{3}{|c|}{SNR=20dB}  &\multicolumn{3}{c|}{SNR=30dB} &\multicolumn{3}{c}{SNR=40dB}\\ \hline
                & linear & bilinear & PNMM & linear & bilinear & PNMM & linear & bilinear & PNMM \\   \hline
     VCA-K-Hype      & 0.0515 & 0.0594 & 0.0443 & 0.0422 & 0.0698 & 0.0443 &0.0362  &0.0384  &0.0436  \\
     VCA-MLM         & 0.0273 & 0.0796 & 0.0360 & 0.0098 & 0.0658 & 0.0438 &0.0090  &0.0560  &0.0299  \\
     N-FINDR-NDU     &0.1186  &0.1141  & 0.0819 & 0.1165 & 0.1112 &0.0747  &0.1140  &0.1069  &0.0718  \\
     rNMF            & 0.0882 & 0.0935 & 0.0746 & 0.0814 & 0.0859 & 0.0710 &0.0814  &0.0816  &0.0716  \\
     NAE             &\textbf{0.0241} & 0.0427 & 0.0373 & 0.0211 & 0.0427 & 0.0372 &0.0189  &0.0200  &0.0368  \\
     NUSAL   & 0.0395 & 0.0434 & 0.0429 & 0.0384 & 0.0430 & 0.0297 &0.0377  &0.0385 &0.0271  \\
     SAE      & 0.0835 & 0.0944 & 0.0953 & 0.0256 & 0.0774 & 0.0464 &0.0208  &0.0460  &0.0660 \\
     Proposed method & \textbf{0.0241} & \textbf{0.0420} & \textbf{0.0304} & \textbf{0.0091} & \textbf{0.0402} & \textbf{0.0292} &\textbf{0.0084}  &\textbf{0.0154}  &\textbf{0.0239} \\
     \hline
     \hline
   \end{tabular}
  \label{tab:result_RMSE}
  \vspace{1mm}
  \\\hspace{-9.2cm} Boldface numbers denote the lowest RMSEs
\end{table*}
\subsection{Experiments with synthetic data}
\subsubsection{Data description}
The synthetic data were generated with the linear mixutre model and two nonlinear models. The endmembers used to generate the data were extracted from USGS digital spectral library. These spectra consist of 224 contiguous bands. The linear mixture model is given by Eq.~\eqref{eq.LMM}. The bilinear mixture model
 \begin{equation}\label{eq.bilinear}
  \mathbf{x}=\mathbf{M}\mathbf{a}+\sum_{i=1}^{R-1}\sum_{j=i+1}^{R}{a}_{i}{a}_{j}\left(\mathbf{m}_{i}\odot \mathbf{m}_{j}\right)+\mathbf{n},
\end{equation}
and  the post-nonlinear mixing model (PPNM):
  \begin{equation}\label{eq.ppnm}
   \mathbf{x}=\mathbf{M}\mathbf{a}+\mathbf{M}\mathbf{a}\odot \mathbf{M}\mathbf{a}+\mathbf{n},
\end{equation}
were used as the two nonlinear models.
In this experiment, four pure material spectra ($R=4$) were considered and the abundance fractions were generated from Dirichlet distribution. A total number of $3\times10^{5}$ pixels were generated to evaluate the performance. Zero-mean Gaussian noise was added with the signal-to-noise ratio (SNR) set to 20 dB, 30 dB and 40 dB, respectively.
Our proposed scheme was implemented using PyTorch and Torch. During the learning process, the data was divided into a number of blocks according to the batch size, which was the number of samples used for a forward operation and a backward propagation operation. In an epoch, the number of iteration was equal to the total number of samples divided by batch size. The weights of the network were updated after learning with each batch. We used Adam optimizer to train the network. Adam is a simple and computationally efficient algorithm for gradient-based optimization of stochastic objective functions. The learning rate was a tuning parameter that determines the step size at each iteration while moving toward a minimum of a loss function. The specific parameters are given in the following subsections.

\subsubsection{Results}
\begin{figure}[!t]
  \centering
  \includegraphics[trim = 10mm 10mm 0 10mm, width=8cm]{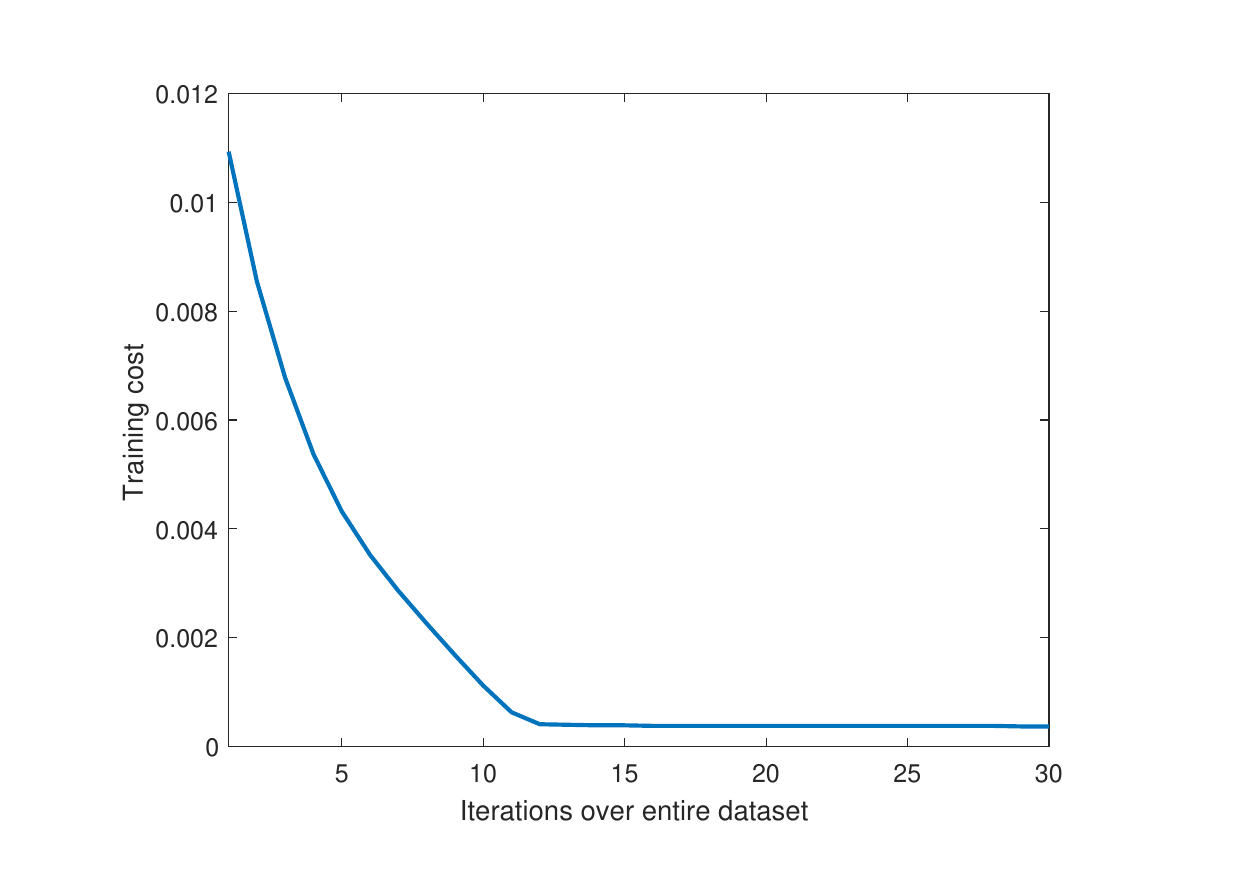}\\
  \caption{Convergence curve during 30 epochs.}\label{fig.loss_iteration}
\end{figure}

\begin{figure}[!t]
  \centering
  \includegraphics[trim = 10mm 10mm 0 10mm, width=6cm]{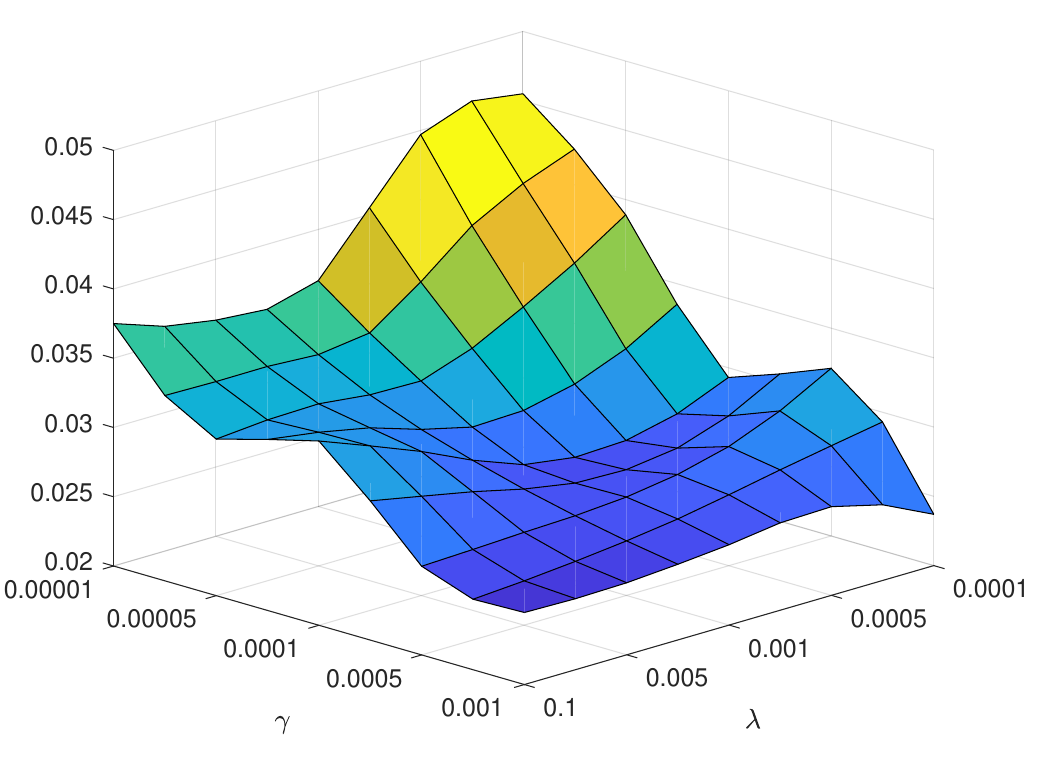}\\
  \caption{RMSE as a function of the regularization parameters for the proposed method.}\label{fig.RMSE_lambda_mu}
\end{figure}

Table~\ref{Tab.network_structure} summarizes the network configurations used in this experiment. The learning rate was set to $1\times10^{-4}$.  The batch size was set to 1024. Note that a larger batch size leads to more accurate descent directions but increases the possibility of reaching a local optimum, while a small batch size may result in difficulties in convergence. The number of training epochs was set to 30. The parameter $\lambda$ was set to $1\times10^{-3}$, and the smoothing regularization parameter $\gamma$ was set to $1\times10^{-3}$. Figure \ref{fig.loss_iteration} shows the convergence curve during learning process with the data generated by the bilinear model under SNR = 30 dB.  We also studied the sensitivity of the proposed method with the algorithm parameters $\lambda$ and $\gamma$ with this data. The result is shown in Figure \ref{fig.RMSE_lambda_mu}. It can be seen the method exhibits satisfactory RMSE within a reasonable range around the optimal parameter values.

Tables~\ref{tab:result_RMSE}, \ref{tab:result_SAD} and \ref{tab:result_SID} report the RMSE, SAD and SID results of the compared methods under different models and SNR settings. It is clear that our proposed method achieves the best abundance estimation performance, and sufficiently good endmember estimation performance with both linear and nonlinear models. Note that when the mixtures are affected by moderate nonlinearities, geometrical endmember extraction algorithms based on linear model can still provide sufficiently good results for nonlinear mixtures, in particular when constraints on simplex volumes are imposed~\cite{dobigeon2014nonlinear}. Among compared algorithms, MLM is a nonlinear unmixing method with a specific assumption on the nonlinearity. Both K-Hype and NDU are kernel-based methods, and the selection of the kernel and its parameters notably affect their performance. The proposed method builds a model by learning the nonlinearity from the observed data, and therefore the issue of the kernel selection is then avoided.
Compared to the state-of-the-art unsupervised nonlinear unmixing methods, namely rNMF and NAE with the same initialization, our proposed method almost always improves the abundance estimation accuracy. Moreover, benefiting from the fact that the low-dimensional vector generated from encoder maintains the main information and gets rid of redundant information and noise, the proposed method is robust to noise.
In order to understand the effect of the smoothing regularization, we  show in Figure~\ref{fig.curves} the extracted endmembers with $\gamma$ set to 0 and $1\times10^{-3}$ in the linear case with SNR = 20 dB. Removing this regularization ($\gamma = 0$) leads to noisy estimated endmember curves. The usefulness of this smoothing effect is clearly illustrated.

\begin{table*} [!htp]
\small \centering
\caption{\small Endmember SAD Comparison of the synthetic data.}
\vspace{-1mm}
  \begin{tabular}{c|c|c|c|c|c|c|c|c|c}
     \hline
     \hline
                &\multicolumn{3}{|c|}{SNR=20dB}  &\multicolumn{3}{c|}{SNR=30dB} &\multicolumn{3}{c}{SNR=40dB}\\   \hline
                & linear & bilinear & PNMM & linear & bilinear & PNMM & linear & bilinear & PNMM \\    \hline
     VCA-K-Hype/MLM  & 1.4615 & 2.4438 & \textbf{5.0424} & 0.6747 & 3.3831 & \textbf{4.9690}   &0.4238  &1.0350  &5.0644  \\
     N-FINDR-NDU     & 5.9772 & 6.2460 & 5.0186 & 2.0760 & 3.1794 & 5.1652   &0.7633  &0.9039  &5.2979  \\
     rNMF            & 9.6746 & 9.4880 & 11.6248& 8.0587 & 7.9005 & 11.5988  &7.7533  &8.1922  &11.6332  \\
     NAE             & \textbf{1.3341} & 2.3112 & 5.4361 & \textbf{0.4415} & 3.1273 & 5.3542   &0.3611  &\textbf{0.9000}  &5.3279  \\
     NUSAL      & 2.2180 & 4.3522 & 7.0544 & 0.7918 & 5.3761 & 5.7092 &0.5370  &2.2496  &5.3339  \\
     SAE        & 2.0868 & 3.3338 & 5.4453 & 0.7376 & 4.6808 & 5.2390 &0.4228  &1.7467 &5.0892  \\
     Proposed method & 1.5544 &\textbf{1.9214} & 5.2590 & 0.5886 & \textbf{3.1201} & 5.0861   &\textbf{0.3434}  &0.9900  &\textbf{5.0036} \\
     \hline
     \hline
   \end{tabular}
  \label{tab:result_SAD}
  \vspace{1mm}
  \\\hspace{-9.8cm} Boldface numbers denote the lowest SADs.
\end{table*}

\begin{table*} [!htp]
\small \centering
\caption{\small Endmember SID Comparison of the synthetic data.}
\vspace{-1mm}
  \begin{tabular}{c|c|c|c|c|c|c|c|c|c}
     \hline
     \hline
                &\multicolumn{3}{|c|}{SNR=20dB}  &\multicolumn{3}{c|}{SNR=30dB} &\multicolumn{3}{c}{SNR=40dB}\\    \hline
                & linear & bilinear & PNMM & linear & bilinear & PNMM & linear & bilinear & PNMM \\    \hline
     VCA-K-Hype/MLM  & 0.0014 & 0.0045 & \textbf{0.0133} & 0.0005 & 0.0087 & 0.0130 &0.0002  &\textbf{0.0006}  &\textbf{0.0140}\\
     N-FINDR-NDU     & 0.0291 & 0.0346 & 0.0112 & 0.0036 & 0.0089 & \textbf{0.0123} &0.0006  &0.0008  &0.0134  \\
     rNMF            & 0.0863 & 0.0875 & 0.1282 & 0.0545 & 0.0529 & 0.1314 &0.0504  &0.0627  &0.1301  \\
     NAE             & \textbf{0.0012} & 0.0037 & 0.0158 & \textbf{0.0001} & \textbf{0.0074} & 0.0154 &\textbf{0.0001}  &0.0007  &0.0153  \\
     NUSAL      & 0.0091 & 0.0096 & 0.0256 & 0.0005 & 0.0136 & 0.0143 &0.0014  &0.0027 &0.0158  \\
     SAE        & 0.0043 & 0.0085 & 0.0157 & 0.0005 & 0.0134 & 0.0137 &0.0002  &0.0014  &0.0141  \\
     Proposed method & 0.0015 & \textbf{0.0022} & 0.0151 & 0.0004 & 0.0076 & 0.0136 & \textbf{0.0001} &\textbf{0.0006} & 0.0148 \\
     \hline
     \hline
   \end{tabular}
  \label{tab:result_SID}
  \vspace{1mm}
  \\\hspace{-9.8cm} Boldface numbers denote the lowest SIDs.
\end{table*}
\begin{figure*}
  \centering
  \centerline{\includegraphics[width=15cm]{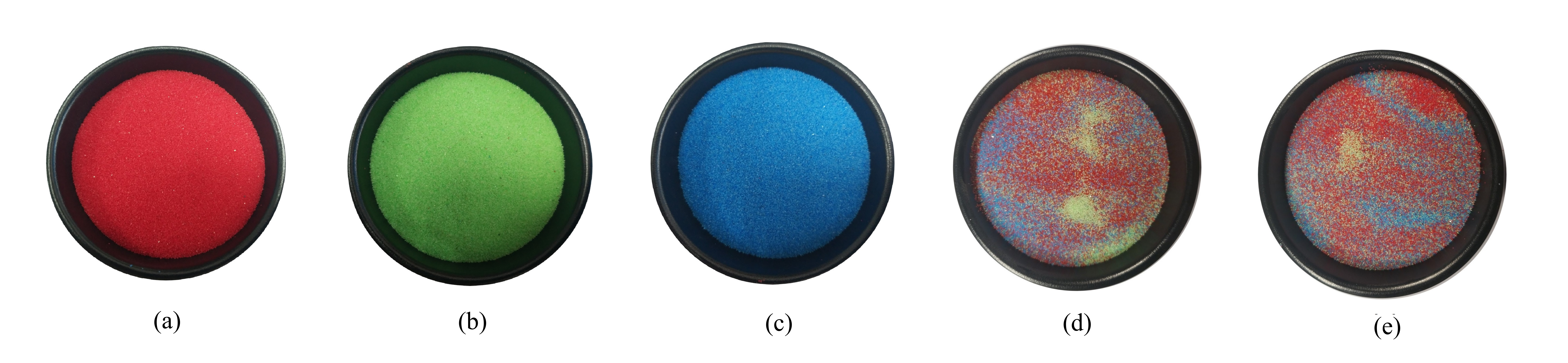}}
  \caption{Laboratory-created data for unmixing performance evaluation (RGB images). Subfigures from (a) to (c): pure quartz sand with the diameter of 0.3 mm of three colors. They serve as pure materials for providing endmembers. Subfigures (d), (e): mixtures of sand with spatial patterns. Square regions of 60-by-60 pixels (0.86 mm$/\text{pixel}$) in the center of each subfigures are clipped out and used in experiments.}
\label{fig.lab_scene}
\end{figure*}

\begin{table*}[t]
 \small \centering
   \caption{RMSE, SAD and SID comparison of unmixing  results of the laboratory-created data.}\label{tab:result_lab}
   \vspace{-1mm}
  \begin{tabular}{c|c|c|c|c|c|c|c|c|c}
     \hline
     \hline
                            &   & VCA-K-Hype & VCA-MLM & N-FINDR-NDU & rNMF & NAE       &NUSAL  &SAE     & Proposed \\\hline
     \multirow{2}{*}{RMSE}  & Mixture 1 & 0.1957 &0.2050 & 0.2135  & 0.2315  &0.2035  &0.2136   &0.2296   & \textbf{0.1942} \\
                            & Mixture 2 & 0.1764 &0.2198 & 0.1961  & 0.2212  &0.1797  &0.2222  &0.2072 & \textbf{0.1729} \\\hline
     \multirow{2}{*}{SAD}   & Mixture 1 & 10.7889&---    & \textbf{9.2097}  & 19.8765 &10.7815  &13.0731  &12.1681 & 10.0852 \\
                            & Mixture 2 & 9.3823 &---    & 12.2489 & 15.5301 &9.3736   &14.3726   &12.3268   & \textbf{9.1427} \\\hline
     \multirow{2}{*}{SID}   & Mixture 1 & 0.0995 &---    & \textbf{0.0325}  & 0.2128  &0.0994   &0.1089   &0.0797  & 0.0986 \\
                            & Mixture 2 & 0.0326 &---    & 0.0568  & 0.1233  &\textbf{0.0325}   &0.1170   &0.0552  & 0.0327 \\
     \hline
     \hline
   \end{tabular}
     \vspace{1mm}
  \\\hspace{-9cm} Boldface numbers denote the lowest RMSEs, SADs and SIDs.
\end{table*}

\begin{figure*}[!t]
  \centering
  \centerline{\includegraphics[width=17.5cm]{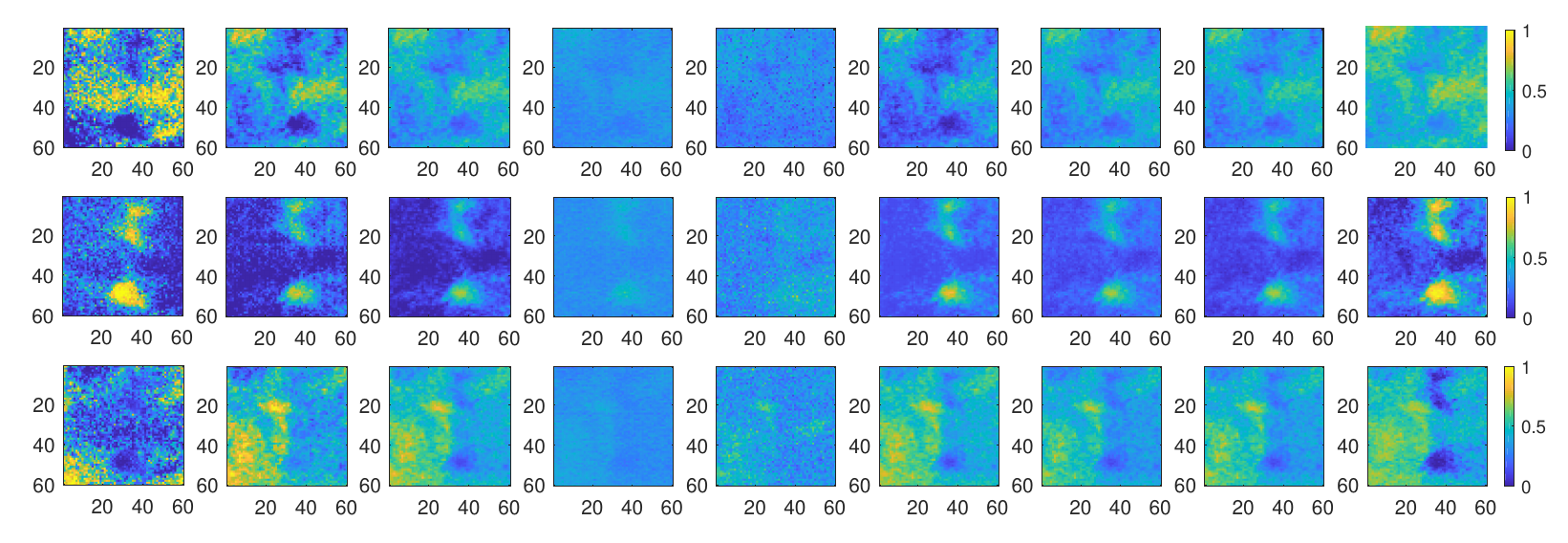}}
  \vspace{-2mm}
  \caption{Abundances maps of the 1st mixture of the laboratory-created data. From left to right columns: ground-truth, estimated results
of VCA-K-Hype, VCA-MLM, N-FINDR-NDU, rNMF, NAE, NUSAL, SAE and the proposed method respectively. From top to bottom: abundance maps of red quartz sand, green quartz sand and blue quartz sand respectively.}
\label{fig.abundance_map_lab_1}
\vspace{-2mm}
\end{figure*}

\begin{figure*}[!t]
  \centering
  \centerline{\includegraphics[width=17.5cm]{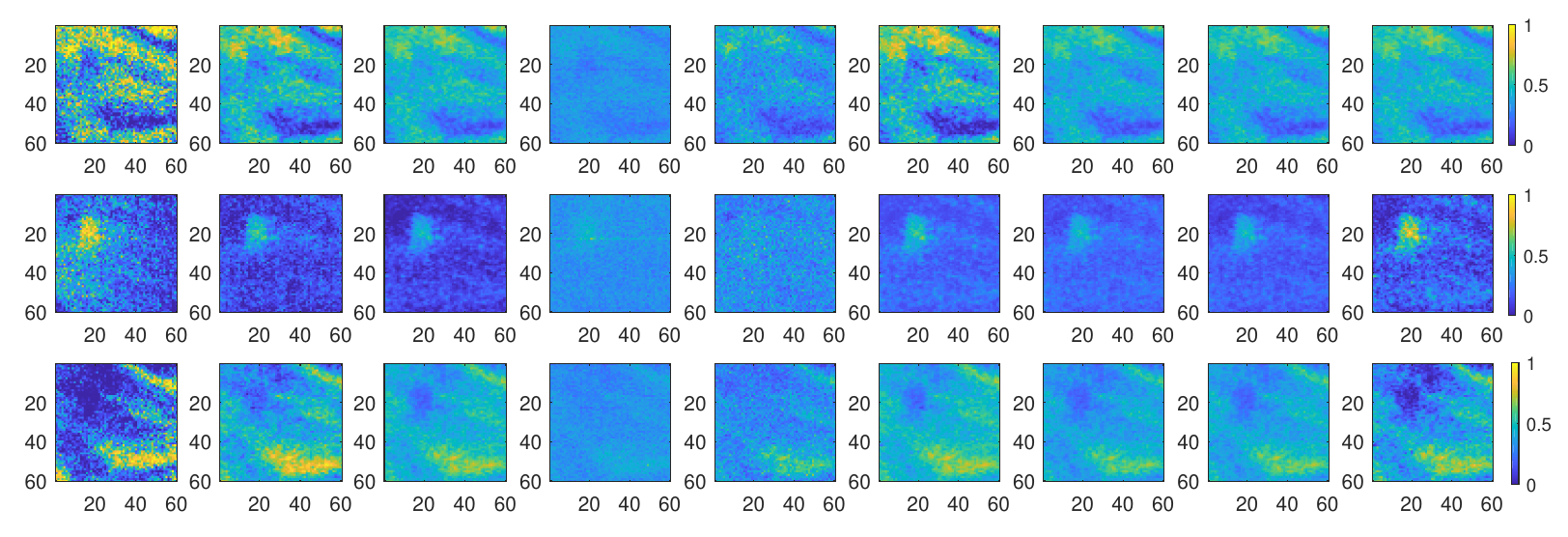}}
  \vspace{-2mm}
  \caption{Abundances maps of the 2nd mixture of the laboratory-created data. From left to right columns: ground-truth, estimated results
of VCA-K-Hype, VCA-MLM, N-FINDR-NDU, rNMF, NAE, NUSAL, SAE and the proposed method respectively. From top to bottom: abundance maps of red quartz sand, green quartz sand and blue quartz sand respectively.}
\label{fig.abundance_map_lab_2}
\vspace{-2mm}
\end{figure*}

\begin{figure}[!t]
  \centering
  \centerline{\includegraphics[width=7cm]{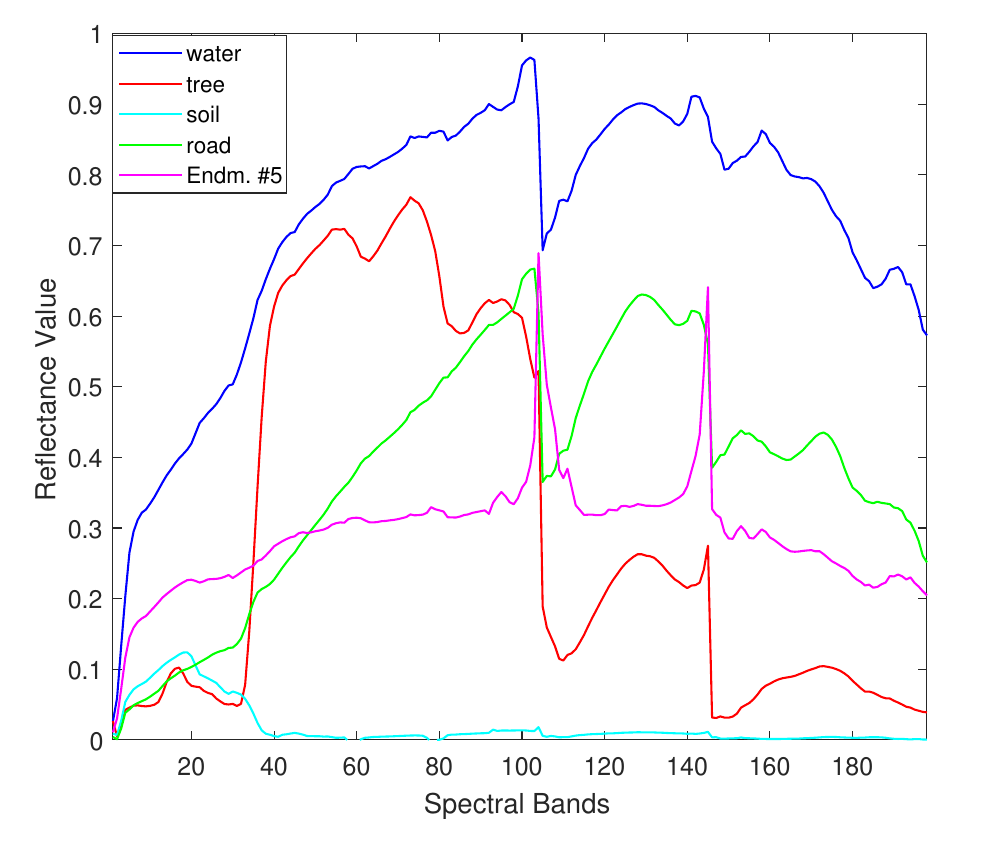}}
  \vspace{-3mm}
  \caption{Extracted endmembers from the Jasper Ridge data by the proposed algorithm.}
\label{fig.curve_jas}
\vspace{-5mm}
\end{figure}


\subsection{Experiments with real data}
\subsubsection{Experiment with laboratory-created data}
In order to perform quantitative evaluation of unmixing performance with real data, we designed several experimental scenes with known ground-truth in our laboratory.
Our data were collected by the GaiaField and GaiaSorter systems in our laboratory. Our GaiaField (Sichuan Dualix Spectral Image Technology Co. Ltd., GaiaField-V10) is a push-broom imaging spectrometer with an HSIA-OL50 lens, covering the visible and NIR wavelengths ranging from 400 nm to 1000 nm, with a spectral resolution up to 0.58 nm.
GaiaSorter sets an environment that isolates external lights, and is endowed with a conveyer to move samples for the push-broom imaging.

Two non-uniform mixtures of colored quartz sand with spatial patterns of Scene-II in our published dataset~\cite{JSTARS2019} were used. The experimental settings were strictly controlled so that pure material spectral signatures and material compositions were known. The data consist of 256 spectral bands.
Different colors of quartz sands with uniform size were used as pure materials shown in Figure~\ref{fig.lab_scene} (a)--(c).
The mixtures are shown in Figure~\ref{fig.lab_scene} (d)--(e).
To calculate the ground-truth, the aligned high-resolution RGB images of these scenes were captured and linked to hyperspectral pixels using the
spatial resolution ratio, and then the percentage of each colored sand in a low-resolution hyperspectral pixel could be analyzed with the help of the associated RGB image. In our experiments, sub-image of 60-by-60 were clipped out from the center of each subfigures.

In this set of experiments, the learning rate was also set to $1\times10^{-4}$, and the Adam optimizer was used to train the network.  The batch size of this experiment was set to 100 and the number of training epochs was set to 50.  The parameter $\lambda$ was set to $1\times10^{-4}$, and $\gamma$ was set to $1\times10^{-6}$.
Figures~\ref{fig.abundance_map_lab_1} and~\ref{fig.abundance_map_lab_2} illustrate the estimated abundance maps of these algorithms. The ground-truth abundance maps are shown in the first columns of Figure~\ref{fig.abundance_map_lab_1} and~\ref{fig.abundance_map_lab_2}. The abundance maps estimated using the compared algorithms and our proposed method are shown alongside. The proposed algorithm results in sharper abundance maps, and the general spatial patterns of the estimated maps are more consistent with the ground-truth. The quantitative RMSE, SAD and SID results are compared in Table~\ref{tab:result_lab}. We observe that the proposed algorithm achieves the lowest RMSEs and sufficiently good endmember estimation performance. These unmixing results with labeled real data highlight the superior performance of the proposed method.

%



\subsubsection{Experiment with real airborne data}
Two real airborne images, namely, Jasper Ridge dataset and urban dataset, were used to validate  the proposed scheme.

\textbf{Jasper Ridge} is a widely used hyperspectral dataset. A subimage of $100\times100$ pixels were used to test the performance of our proposed method and several other compared algorithms. Each pixel was recorded at 224 channels ranging from 380 nm to 2500 nm with spectral resolution up to 9.46 nm.
After removing the channels affected by water vapor and the atmospheric environment, 198 channels were kept. The number of endmembers was set to 5, including water, tree, soil, road, and the 5th endmember.

\begin{figure*}
  \centering
  \centerline{\includegraphics[width=19cm]{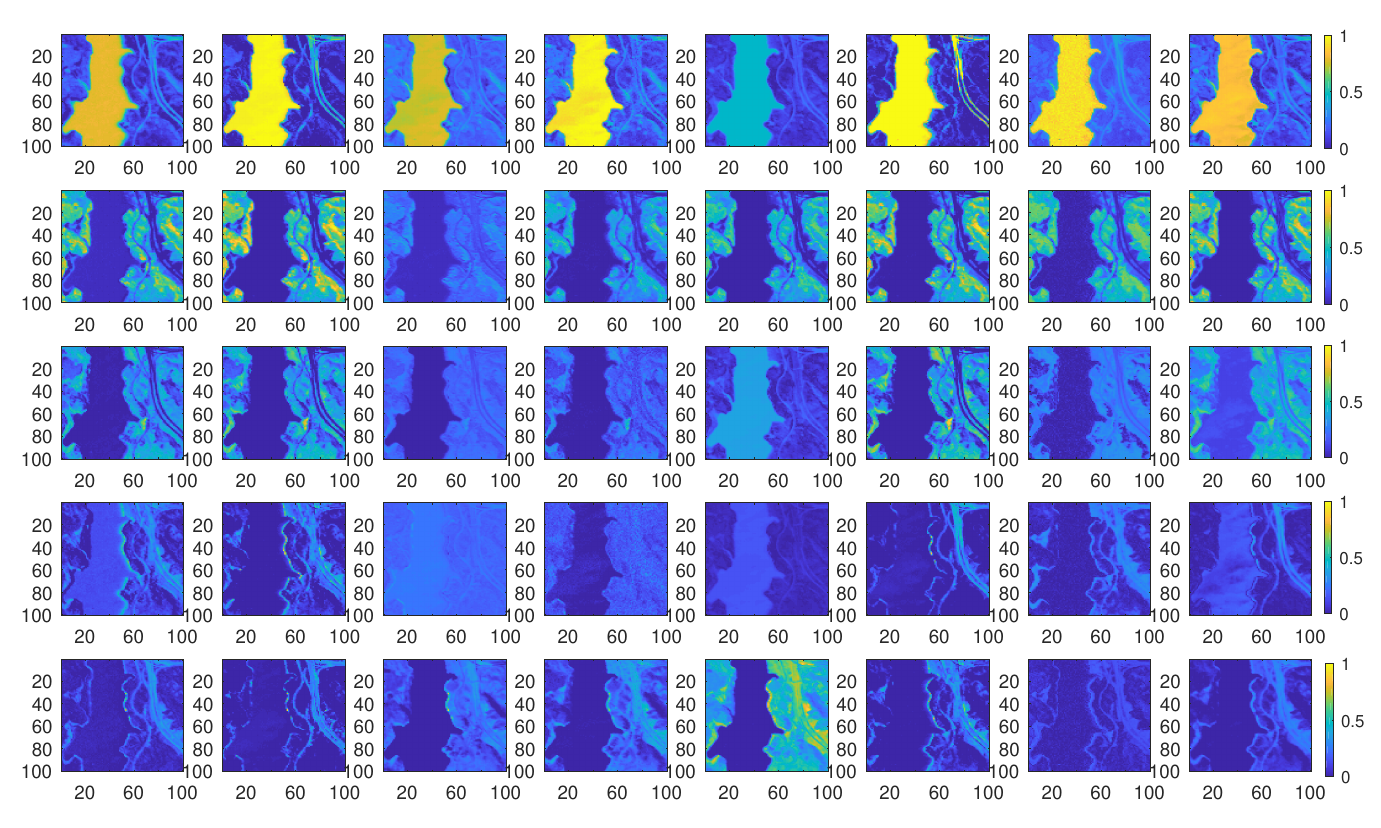}}
  \vspace{-4mm}
  \caption{Estimated abundance maps of Jasper Ridge data. From left to right: VCA-K-Hype, VCA-MLM, N-FINDR-NDU, rNMF, NAE, NUSAL, SAE and the proposed method. From top to bottom: water, tree, soil, road, and the 5th  endmember respectively.}
\label{fig.map_jas}
\end{figure*}

\begin{figure*}
  \centering
  \centerline{\includegraphics[width=18cm]{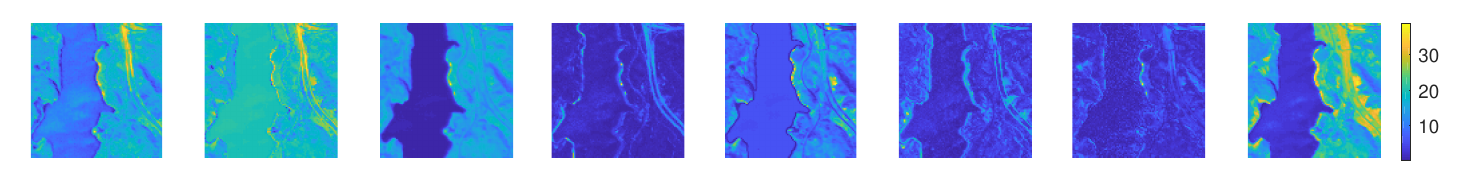}}
  \vspace{-4mm}
  \caption{Energy of the nonlinear components of the Jasper Ridge data. From left to right: VCA-K-Hype, VCA-MLM, N-FINDR-NDU, rNMF, NAE, NUSAL, SAE and the proposed method.}
\label{fig.map_nonlinear_jas}
\end{figure*}

The same network defined in Table~\ref{Tab.network_structure} was used for this data, with $B=198$, and $R=5$, with the learning rate set to $1\times10^{-4}$. The batch size used in this experiment was set to 512, and the number of training epochs was set to 50. The parameter $\lambda$ was set to $1\times10^{-3}$, and $\gamma$ was set to $1\times10^{-8}$ in this experiment. Figure~\ref{fig.curve_jas} illustrates the extracted endmembers by the proposed algorithm. Figure~\ref{fig.map_jas} illustrates the estimated abundance maps of the five endmembers obtained by these algorithms.
We observe that the proposed algorithm provides a shaper and clearer map of different materials. Figure~\ref{fig.map_nonlinear_jas} shows the energy of the nonlinear components estimated by these algorithms, and the nonlinear energy of the pixel is the sum of the nonlinear component of all spectral bands. These maps demonstrate that nonlinear components are active at the boundary or transition parts of different regions, e.g. at the water shore. The proposed algorithm provides a clear map of nonlinear components with several particular locations emphasized.

\begin{figure*}[!t]
  \centering
  \centerline{\includegraphics[width=18.5cm]{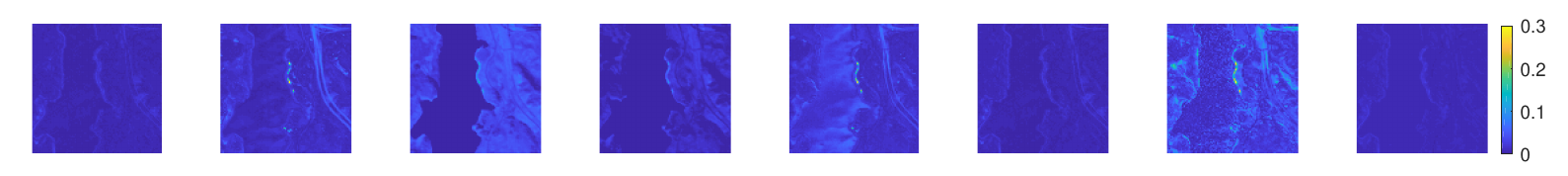}}
  \vspace{-4mm}
  \caption{Maps of reconstruction error of the Jasper Ridge data. From left to right: VCA-K-Hype, VCA-MLM, N-FINDR-NDU, rNMF, NAE, NUSAL, SAE and the proposed method respectively.}
\label{fig.map_RE_jas}
\end{figure*}

Note that this real data is extensively used in hyperspectral unmixing,
however, no ground-truth information is available for a quantitative performance evaluation of abundance. Thus, the reconstruction error (RE) defined by:
\begin{equation}\label{eq.RE}
  RE=\sqrt{\frac{1}{NR}\sum_{i=1}^{N}\|\mathbf{x}_{i}-\mathbf{\widehat{x}}_{i}\|_2}
\end{equation}
is used for a quantitative comparison, where $\mathbf{x}_i$ and $\mathbf{\widehat{x}}_{i}$ denote the true and reconstructed vector of the $i$th pixel, $N$ represents the total number of pixels, though RE may not be proportional to the abundance estimation accuracy. The RE results of different algorithms are reported in Table~\ref{tab:RE_result_jas}, and the reconstruction error maps are illustrated in Figure~\ref{fig.map_RE_jas}. We observe that our method leads to the lowest reconstruction error in the mean sense and in the spatial distribution.


\begin{table*}[!t]
 \small \centering
   \caption{RE comparison of the Jasper Ridge data.}\label{tab:RE_result_jas}
   \vspace{-1mm}
  \begin{tabular}{ccccccccc}
     \hline
     \hline
     Algorithm & VCA-K-Hype & VCA-MLM & N-FINDR-NDU & rNMF    & NAE    & NUSAL  &SAE   & Proposed \\
     RE        & 0.0128     &0.0250   & 0.0866      & 0.0517  &0.0298  &0.0114  &0.0937    &\textbf{0.0111}  \\
    \hline
     \hline
   \end{tabular}
   \vspace{1mm}
     \\\hspace{-7.8cm} Boldface numbers denote the lowest RE value.
\end{table*}

\begin{table*}
 \small \centering
   \caption{RE comparison of the Urban data.}\label{tab:RE_result}
   \vspace{-1mm}
  \begin{tabular}{ccccccccc}
     \hline
     \hline
     Algorithm & VCA-K-Hype & VCA-MLM & N-FINDR-NDU & rNMF  & NAE    & NUSAL  & SAE  & Proposed \\
     RE        & 0.0194    &0.0139   & 0.0241    & 0.0176     &0.0211  &0.0123  &0.0154   &\textbf{0.0113}  \\

     \hline
     \hline
   \end{tabular}
      \vspace{1mm}
    \\\hspace{-7.8cm} Boldface numbers denote the lowest RE value.
\end{table*}

\textbf{Urban} is a widely used hyperspectral dataset for unmixing task. The urban image\footnote{http://www.agc.army.mil/hypercube/} has $307\times 307$ pixels. All the pixels were used to evaluate the unmixing performance. The data consist of 210 spectral bands ranging from 400 nm to 2500 nm with spectral resolution up to 10 nm. After removing channels [1-4, 76, 87, 101-111, 136-153, 198-210] affected by dense water vapor and the atmosphere, 162 channels were remained. Five prominent endmembers exist in this data, namely, asphalt, grass, tree, roof, and dirt.

\begin{figure}[!t]
  \centering
  \centerline{\includegraphics[width=8cm]{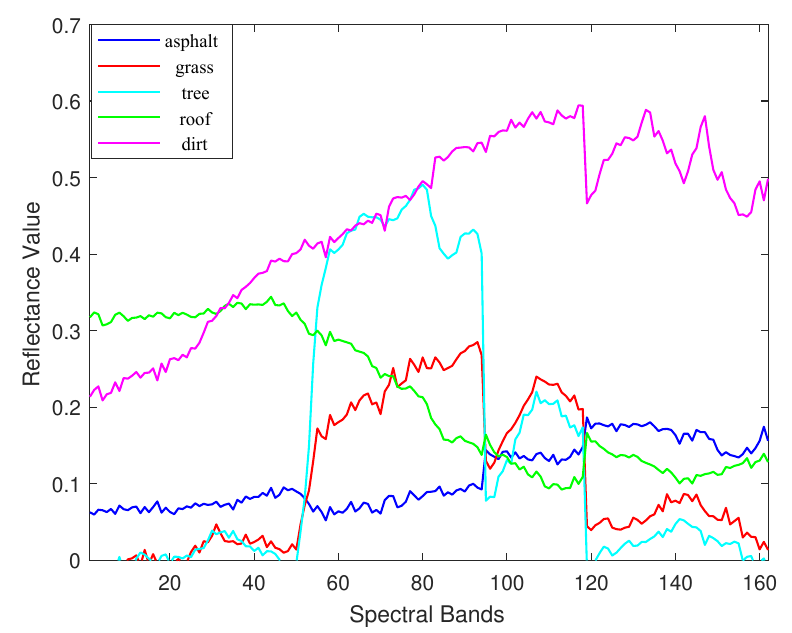}}
  \caption{Extracted endmembers from the urban data by the proposed algorithm.}
\label{fig.curve_urban}
\vspace{-2mm}
\end{figure}

In this experiment, the same network, learning rate, and optimizer were used to conduct the unmixing study. The batch size was set to 512, with the number of epoch set to 80. The parameter $\lambda$ was set to $1\times 10^{-4}$ and $\gamma$ was set to $1\times 10^{-6}$.  Figure~\ref{fig.curve_urban} shows the extracted endmembers by the proposed method.

\begin{figure*}
  \centering
  \centerline{\includegraphics[width=19cm]{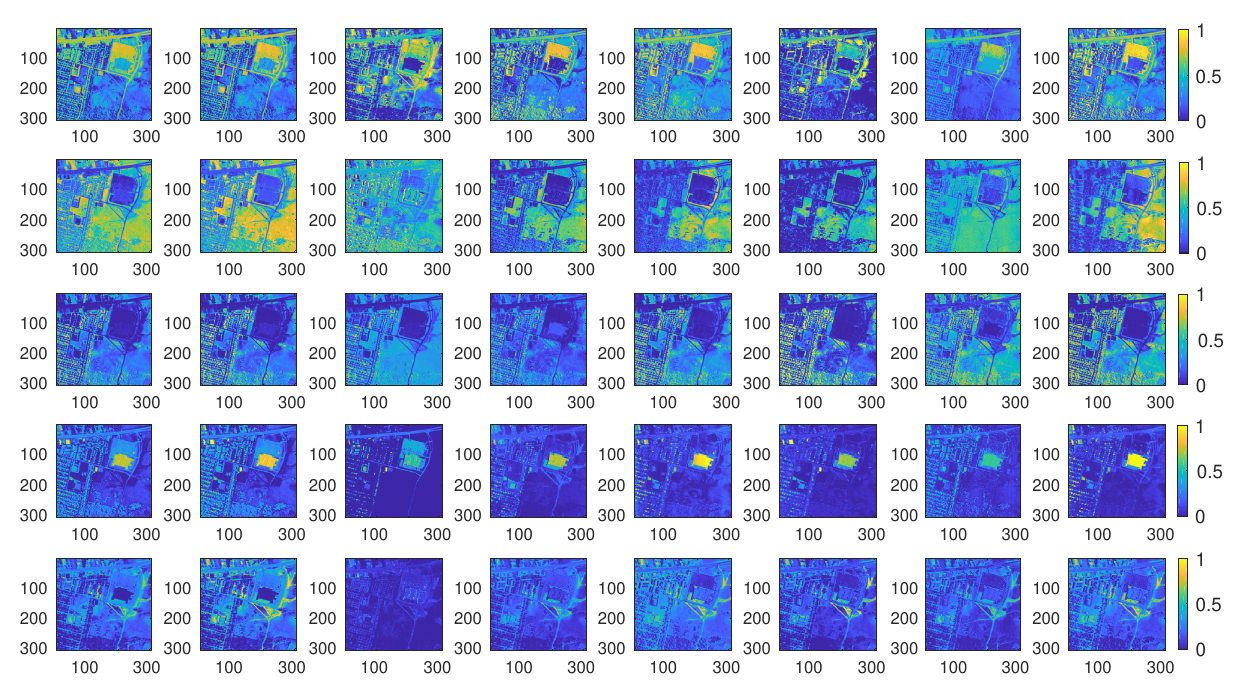}}
  \caption{Estimated abundance maps of urban data. From left to right: VCA-K-Hype, VCA-MLM, N-FINDR-NDU, rNMF, NAE, NUSAL, SAE and the proposed method. From top to bottom: asphalt, grass, tree, roof, dirt.}
\label{fig.map_urban}
\end{figure*}

\begin{figure*}
  \centering
  \centerline{\includegraphics[width=18cm]{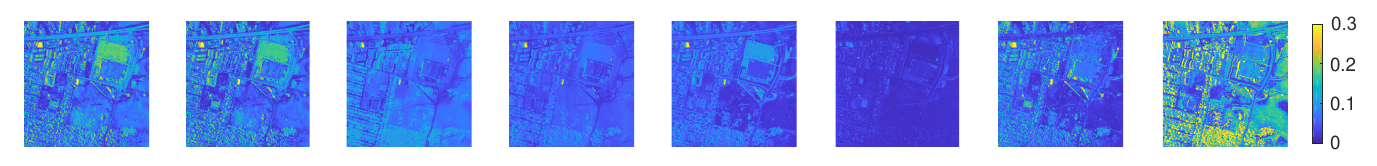}}
  \caption{Energy of the nonlinear components of the urban data. From left to right: VCA-K-Hype, VCA-MLM, N-FINDR-NDU, rNMF, NAE, NUSAL, SAE and the proposed method.}
\label{fig.map_nonlinear}
\vspace{-2mm}
\end{figure*}

\begin{figure*}
  \centering
  \centerline{\includegraphics[width=18.5cm]{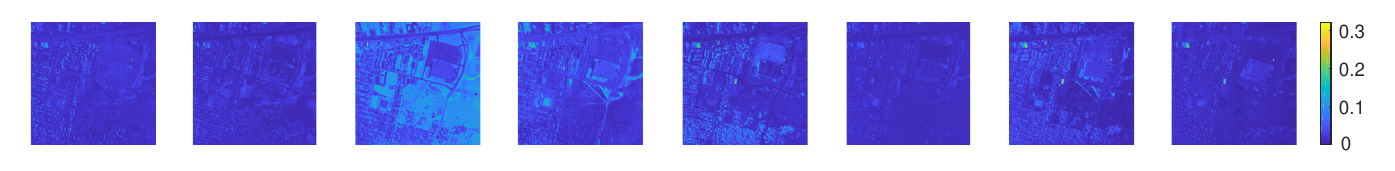}}
  \caption{Maps of reconstruction error the urban data. From left to right: VCA-K-Hype, VCA-MLM, N-FINDR-NDU, rNMF, NAE, NUSAL, SAE and the proposed method.}
\label{fig.RE_urban}
\vspace{-2mm}
\end{figure*}

The estimated abundance maps of five endmembers are shown in Figure~\ref{fig.map_urban}.
These figures clearly indicate that our proposed method provides a smoother and clearer map.
Figure~\ref{fig.map_nonlinear} shows the energy of the nonlinear components estimated by these algorithms. These maps demonstrate that nonlinear components are active at vegetated regions and boundary or transition parts of different regions. The proposed algorithm provides a clearer map of nonlinear components.
The RE results achieved by different algorithms are reported in Table \ref{tab:RE_result}, and the reconstructed error maps are shown in Figure \ref{fig.RE_urban}. We observed that our method leads to the lowest reconstruction error.

\section{Conclusion}
This paper presented an unsupervised nonlinear spectral unmixing method based on a deep autoencoder network that applied a general mixture model consisting of a linear mixture component and an additive nonlinear mixture component. The proposed approach benefits from the universal modeling ability of deep neural networks to learn the inherent nonlinearity of the nonlinear mixture component from the data itself via the autoencoder network.
The superior performance of the proposed method was validated with both synthetic and real data, particularly with laboratory-created labeled data. Future work will integrate contexture information of the image into the autoencoder network and consider the spectral variability problem to further enhance its performance.

\bibliographystyle{IEEEtran}
\bibliography{IEEEfull,BIB}\ 

\begin{thebibliography}{10}
\providecommand{\url}[1]{#1}
\csname url@samestyle\endcsname
\providecommand{\newblock}{\relax}
\providecommand{\bibinfo}[2]{#2}
\providecommand{\BIBentrySTDinterwordspacing}{\spaceskip=0pt\relax}
\providecommand{\BIBentryALTinterwordstretchfactor}{4}
\providecommand{\BIBentryALTinterwordspacing}{\spaceskip=\fontdimen2\font plus
\BIBentryALTinterwordstretchfactor\fontdimen3\font minus
  \fontdimen4\font\relax}
\providecommand{\BIBforeignlanguage}[2]{{%
\expandafter\ifx\csname l@#1\endcsname\relax
\typeout{** WARNING: IEEEtran.bst: No hyphenation pattern has been}%
\typeout{** loaded for the language `#1'. Using the pattern for}%
\typeout{** the default language instead.}%
\else
\language=\csname l@#1\endcsname
\fi
#2}}
\providecommand{\BIBdecl}{\relax}
\BIBdecl

\bibitem{bioucas2012hyperspectral}
J.~M. Bioucas-Dias, A.~Plaza, N.~Dobigeon, M.~Parente, Q.~Du, P.~Gader, and
  J.~Chanussot, ``Hyperspectral unmixing overview: Geometrical, statistical,
  and sparse regression-based approaches,'' \emph{IEEE J. Sel. Top. Appl. Earth
  Observat. Remote Sens.}, vol.~5, no.~2, pp. 354--379, 2012.

\bibitem{Keshava2002Spectral}
N.~Keshava and J.~F. Mustard, ``Spectral unmixing,'' \emph{IEEE Signal Proc.
  Mag.}, vol.~19, no.~1, pp. 44--57, 2002.

\bibitem{yao2019nonconvex}
J.~Yao, D.~Meng, Q.~Zhao, W.~Cao, and Z.~Xu, ``Nonconvex-sparsity and
  nonlocal-smoothness-based blind hyperspectral unmixing,'' \emph{IEEE Trans.
  Image Process.}, vol.~28, no.~6, pp. 2991--3006, 2019.

\bibitem{hong2018augmented}
D.~Hong, N.~Yokoya, J.~Chanussot, and X.~Zhu, ``An augmented linear mixing
  model to address spectral variability for hyperspectral unmixing,''
  \emph{IEEE Trans. Image Process.}, vol.~28, no.~4, pp. 1923--1938, 2018.

\bibitem{Heylen2014A}
R.~Heylen, M.~Parente, and P.~Gader, ``A review of nonlinear hyperspectral
  unmixing methods,'' \emph{IEEE J. Sel. Top. Appl. Earth Observat. Remote
  Sens.}, vol.~7, no.~6, pp. 1844--1868, 2014.

\bibitem{Borel1994Nonlinear}
C.~Borel and S.~Gerstl, ``Nonlinear spectral mixing models for vegetative and
  soil surfaces,'' \emph{Remote Sens. Envi.}, vol.~47, no.~3, pp. 403--416,
  1994.

\bibitem{Fan2009Comparative}
W.~Fan, B.~Hu, J.~Miller, and M.~Li, ``Comparative study between a new
  nonlinear model and common linear model for analysing laboratory
  simulated-forest hyperspectral data,'' \emph{Int. J. Remote Sens.}, vol.~30,
  no.~11, pp. 2951--2962, 2009.

\bibitem{Halimi2011Unmixing}
A.~Halimi, Y.~Altmann, N.~Dobigeon, and J.~Tourneret, ``Unmixing hyperspectral
  images using the generalized bilinear model,'' in \emph{Geoscience Remote
  Sensing Symposium}, 2011.

\bibitem{Altmann2012Supervised}
Y.~Altmann, A.~Halimi, N.~Dobigeon, and J.~Tourneret, ``Supervised nonlinear
  spectral unmixing using a postnonlinear mixing model for hyperspectral
  imagery,'' \emph{IEEE Trans. Image Process.}, vol.~21, no.~6, pp. 3017--3025,
  2012.

\bibitem{Hapke1981Bidirectional}
B.~Hapke, ``Bidirectional reflectance spectroscopy: 1. theory,'' \emph{J.
  Geophys. Res.}, vol.~86, no.~B4, pp. 3039--3054, 1981.

\bibitem{Close2012Using}
R.~Close, P.~Gader, J.~Wilson, and A.~Zare, ``Using physics-based macroscopic
  and microscopic mixture models for hyperspectral pixel unmixing,''
  \emph{Proc. Soc. Photo-Opt. Instrum. Eng. (SPIE), Algorithms Technol.
  Multispectral Hyperspectral Ultraspectral Imagery XVIII}, vol. 8390, no.~1,
  pp. 83\,901L--1, 2012.

\bibitem{chen2013nonlinear}
J.~Chen, C.~Richard, and P.~Honeine, ``Nonlinear unmixing of hyperspectral data
  based on a linear-mixture/nonlinear-fluctuation model,'' \emph{IEEE Trans.
  Signal Process.}, vol.~61, no.~2, pp. 480--492, 2013.

\bibitem{Chen2014TGRS}
------, ``Nonlinear estimation of material abundances in hyperspectral images
  with $\ell_{1}$-norm spatial regularization,'' \emph{IEEE Trans. Geosci.
  Remote Sens.}, vol.~52, no.~5, pp. 2654--2665, 2014.

\bibitem{ammanouil2017nonlinear}
R.~Ammanouil, A.~Ferrari, C.~Richard, and S.~Mathieu, ``Nonlinear unmixing of
  hyperspectral data with vector-valued kernel functions,'' \emph{IEEE Trans.
  Image Process.}, vol.~26, no.~1, pp. 340--354, 2017.

\bibitem{heylen2016multilinear}
R.~Heylen and P.~Scheunders, ``A multilinear mixing model for nonlinear
  spectral unmixing,'' \emph{IEEE Trans. Geosci. Remote Sens.}, vol.~54, no.~1,
  pp. 240--251, 2016.

\bibitem{heylen2019nonlinear}
R.~Heylen, V.~Andrejchenko, Z.~Zahiri, M.~Parente, and P.~Scheunders,
  ``Nonlinear hyperspectral unmixing with graphical models,'' \emph{IEEE Trans.
  Geosci. Remote Sens.}, vol.~57, no.~7, pp. 4844--4856, 2019.

\bibitem{chen2014deep}
Y.~Chen, Z.~Lin, X.~Zhao, G.~Wang, and Y.~Gu, ``Deep learning-based
  classification of hyperspectral data,'' \emph{IEEE J. Sel. Top. Appl. Earth
  Observat. Remote Sens.}, vol.~7, no.~6, pp. 2094--2107, 2014.

\bibitem{mou2017deep}
L.~Mou, P.~Ghamisi, and X.~Zhu, ``Deep recurrent neural networks for
  hyperspectral image classification,'' \emph{IEEE Trans. Geosci. Remote
  Sens.}, vol.~55, no.~7, pp. 3639--3655, 2017.

\bibitem{chen2015spectral}
Y.~Chen, X.~Zhao, and X.~Jia, ``Spectral--spatial classification of
  hyperspectral data based on deep belief network,'' \emph{IEEE J. Sel. Top.
  Appl. Earth Observat. Remote Sens.}, vol.~8, no.~6, pp. 2381--2392, 2015.

\bibitem{licciardi2011pixel}
G.~Licciardi and F.~Del~Frate, ``Pixel unmixing in hyperspectral data by means
  of neural networks,'' \emph{IEEE Trans. Geosci. Remote Sens.}, vol.~49,
  no.~11, pp. 4163--4172, 2011.

\bibitem{zhang2018hyperspectral}
X.~Zhang, Y.~Sun, J.~Zhang, P.~Wu, and L.~Jiao, ``Hyperspectral unmixing via
  deep convolutional neural networks,'' \emph{IEEE Geosci. Remote Sens. Lett},
  no.~99, pp. 1--5, 2018.

\bibitem{1guo2015hyperspectral}
R.~Guo, W.~Wang, and H.~Qi, ``Hyperspectral image unmixing using autoencoder
  cascade,'' in \emph{WHISPERS 2015}.\hskip 1em plus 0.5em minus 0.4em\relax
  IEEE, 2015, pp. 1--4.

\bibitem{1palsson2018hyperspectral}
B.~Palsson, J.~Sigurdsson, J.~Sveinsson, and M.~Ulfarsson, ``Hyperspectral
  unmixing using a neural network autoencoder,'' \emph{IEEE Access}, vol.~6,
  pp. 25\,646--25\,656, 2018.

\bibitem{1qu2017spectral}
Y.~Qu, R.~Guo, and H.~Qi, ``Spectral unmixing through part-based non-negative
  constraint denoising autoencoder,'' in \emph{Proc. IEEE International
  Geoscience and Remote Sensing Symposium (IGARSS)}, 2017, pp. 209--212.

\bibitem{1qu2018udas}
Y.~Qu and H.~Qi, ``{uDAS}: An untied denoising autoencoder with sparsity for
  spectral unmixing,'' \emph{IEEE Trans. Geosci. Remote Sens.}, vol.~57, no.~3,
  pp. 1698--1712, 2018.

\bibitem{1su2017nonnegative}
Y.~Su, A.~Marinoni, J.~Li, A.~Plaza, and P.~Gamba, ``Nonnegative sparse
  autoencoder for robust endmember extraction from remotely sensed
  hyperspectral images,'' in \emph{Proc. IEEE International Geoscience and
  Remote Sensing Symposium (IGARSS)}, 2017, pp. 205--208.

\bibitem{1su2018stacked}
Y.~Su, J.~Li, A.~Plaza, A.~Marinoni, P.~Gamba, and S.~Chakravortty, ``{DAEN}:
  Deep autoencoder networks for hyperspectral unmixing,'' \emph{IEEE Trans.
  Geosci. Remote Sens.}, vol.~57, no.~7, pp. 4309--4321, 2019.

\bibitem{ozkan2018endnet}
S.~Ozkan, B.~Kaya, and G.~B. Akar, ``{EndNet}: Sparse autoencoder network for
  endmember extraction and hyperspectral unmixing,'' \emph{IEEE Trans. Geosci.
  Remote Sens.}, vol.~57, no.~1, pp. 482--496, 2018.

\bibitem{borsoi2019deep}
R.~Borsoi, T.~Imbiriba, and J.~Bermudez, ``Deep generative endmember modeling:
  An application to unsupervised spectral unmixing,'' \emph{IEEE Trans. Comput.
  Imag.}, pp. 1--1, 2019.

\bibitem{NAE2019}
M.~Wang, M.~Zhao, J.~Chen, and S.~Rahardja, \emph{IEEE Geosci. Remote Sens.
  Lett}, vol.~16, no.~9, pp. 1467--1471, 2019.

\bibitem{goodfellow2016deep}
I.~Goodfellow, Y.~Bengio, and A.~Courville, \emph{Deep learning}.\hskip 1em
  plus 0.5em minus 0.4em\relax MIT press, 2016.

\bibitem{winter1999n}
M.~E. Winter, ``{N-FINDR}: An algorithm for fast autonomous spectral end-member
  determination in hyperspectral data,'' in \emph{Imaging Spectrometry V}, vol.
  3753.\hskip 1em plus 0.5em minus 0.4em\relax International Society for Optics
  and Photonics, 1999, pp. 266--276.

\bibitem{F2015Nonlinear}
C.~F{\'e}votte and N.~Dobigeon, ``Nonlinear hyperspectral unmixing with robust
  nonnegative matrix factorization.'' \emph{IEEE Trans. Image Process.},
  vol.~24, no.~12, pp. 4810--4819, 2015.

\bibitem{halimi2016fast}
A.~Halimi, J.~Bioucas-Dias, N.~Dobigeon, G.~Buller, and S.~McLaughlin, ``Fast
  hyperspectral unmixing in presence of nonlinearity or mismodeling effects,''
  \emph{IEEE Trans. Computational Imaging}, vol.~3, no.~2, pp. 146--159, 2016.

\bibitem{xu2018supervised}
X.~Xu, Z.~Shi, and B.~Pan, ``A supervised abundance estimation method for
  hyperspectral unmixing,'' \emph{Remote sens. lett.}, vol.~9, no.~4, pp.
  383--392, 2018.

\bibitem{dobigeon2014nonlinear}
N.~Dobigeon, J.~Tourneret, C.~Richard, J.~Bermudez, S.~McLaughlin, and A.~Hero,
  ``Nonlinear unmixing of hyperspectral images: Models and algorithms,''
  \emph{IEEE Signal Process. Mag.}, vol.~31, no.~1, pp. 82--94, 2014.

\bibitem{JSTARS2019}
M.~Zhao, J.~Chen, and Z.~He, ``A labortary-created dataset with ground-truth
  for hyperspectral unmixing evaluation,'' \emph{IEEE J. Sel. Top. Appl. Earth
  Observat. Remote Sens.}, vol.~12, no.~7, pp. 2170--2183, 2019.

\end{thebibliography}

\end{document}